\documentclass[aps,prc,reprint,floatfix,amsmath,amssymb,longbibliography]{revtex4-2}
\usepackage[T1]{fontenc}
\usepackage{booktabs,graphicx,hyperref,amsthm,pgfplots}
\pgfplotsset{compat=1.18}
\usepgfplotslibrary{groupplots}
\hypersetup{colorlinks=true,citecolor=blue,linkcolor=blue,urlcolor=blue}
\newtheorem{theorem}{Theorem}

\usepackage{graphicx}
\usepackage{bm}

\begin{document}

\title{Higher-order strongly intensive observables for forward-backward multiplicity correlations}%

\author{Evgeny Andronov}
 \email{Contact author: e.v.andronov@spbu.ru}
\affiliation{%
 St. Petersburg State University, 7/9 Universitetskaya nab., St. Petersburg, 199034
Russia
}%

\date{\today}

\begin{abstract}
Forward-backward multiplicity correlations probe the longitudinal structure of particle production, but fluctuations in the number of emitting sources complicate their interpretation. Strongly intensive observables remove the contribution of source-number fluctuations at second order. Here we extend the construction to all cumulant orders. For independent identical sources whose emission law does not depend on the source count, we build a normalized hierarchy of strongly intensive observables with $n$ new algebraically independent quantities at order $n$. Every observable is normalized to unity for statistically independent particles, vanishes when both window multiplicities are fixed, and remains finite at equal positive window means. We give explicit expressions through fourth order and an all-order recursion. Analytical particle production models and PYTHIA calculations for $pp$ collisions at $\sqrt{s}=0.9$ and 13 TeV demonstrate the sensitivity to window geometry, event selection, and connected multiparticle correlations. Fits constrained by the mean multiplicities and second-order observables provide conditional higher-order benchmarks. Deviations of these benchmarks from the PYTHIA results test the assumed correlation hierarchy, subject to the quality of the lower-order fit. Preliminary ALICE measurements at $\sqrt{s}=13$ TeV are used to illustrate the corresponding model-dependent scenarios.
\end{abstract}

\maketitle
\raggedbottom

\section{\label{sec:intro}Introduction}
Rapidity correlations constrain the longitudinal structure of particle production beyond the information contained in the inclusive multiplicity density. Forward-backward analyses measure multiplicities of charged particles in two separated rapidity or pseudorapidity windows. Varying their separation probes the correlation range, whereas varying their widths probes the accumulation of correlations within a finite acceptance. An additional azimuthal separation resolves structure that a full-azimuth analysis integrates over \cite{ALICEFB2015,Vechernin2015FB}. Resonance decays, jets, and fragmentation contribute to short-range correlations. Correlations over larger rapidity intervals are sensitive to longitudinally extended sources, initial-state fluctuations, and fluctuations in the number of particle-producing interactions. These mechanisms have been investigated in string models, wounded-constituent models, early-stage color-field descriptions, and event generators \cite{Amelin:1994mf,Braun:2000cc,Kovalenko:2012ixp,Bzdak2009FB,Konchakovski2009FB,LappiMcLerran2010,Mondal2023FB,SinghHYDJET2026}.

A conventional measure is the forward-backward (FB) multiplicity correlation coefficient. For multiplicities $n_F$ and $n_B$ in two nonoverlapping windows it is defined as
\begin{equation}
b_{\rm corr}=\frac{\operatorname{cov}(n_F,n_B)}{\operatorname{Var}(n_F)}
=\frac{\langle n_Fn_B\rangle-\langle n_F\rangle\langle n_B\rangle}
{\langle n_F^2\rangle-\langle n_F\rangle^2}.
\label{eq:intro_bcorr}
\end{equation}
Here and below brackets denote an average over the selected events. This quantity is the slope of the linear regression of $n_B$ on $n_F$. Its definition is asymmetric under interchange of the windows. It coincides with the Pearson coefficient when the two variances are equal. FB multiplicity correlations have been measured by UA5 in $\bar{p}p$ collisions \cite{UA5FB1988}, by STAR in  $pp$ and Au+Au collisions at $\sqrt{s_{\rm NN}}=200$~GeV \cite{STARFB2009}, by ALICE in  $pp$ collisions at $\sqrt{s}=0.9$, 2.76, and 7~TeV \cite{ALICEFB2015}, and by ATLAS in  $pp$ collisions at 0.9 and 7~TeV \cite{ATLASFB2012}.

An independent-source model provides a useful baseline for interpreting these measurements. Consider $S$ independent identical sources, each emitting multiplicities $\mu_F$ and $\mu_B$ in observation windows. The source emission law is assumed to be independent of $S$. Writing the scaled variance as $\omega[X]=\operatorname{Var}(X)/\langle X\rangle$, the laws of total variance and covariance give \cite{Vechernin2015FB}
\begin{equation}
 b_{\rm corr}=
 \frac{\operatorname{cov}(\mu_F,\mu_B)
    +\langle\mu_F\rangle\langle\mu_B\rangle\omega[S]}
   {\operatorname{Var}(\mu_F)+\langle\mu_F\rangle^2\omega[S]}.
 \label{eq:intro_bsources}
\end{equation}
In Ref.~\cite{Vechernin2015FB} this expression was related to the string pair correlation function. In particular, for two equal narrow pseudorapidity windows of width $\delta\eta$, full azimuth, and a uniform single-string density $\mu_0$, the result becomes
\begin{equation}
 b_{\rm corr}\simeq
 \frac{\mu_0\delta\eta\,[\omega[S]+\Lambda(\Delta\eta)]}
 {1+\mu_0\delta\eta\,[\omega[S]+\Lambda(0)]}.
 \label{eq:intro_bstring}
\end{equation}
Here $\Delta\eta$ is the distance between window centers and $\Lambda$ is the normalized two-particle correlation function of one string, averaged over azimuth. Thus even when the same-string correlation between the windows vanishes, $b_{\rm corr}$ retains a contribution from $\omega[S]$. The energy dependence of this contribution mixes changes in string fragmentation with changes in the distribution of the number of strings. Centrality selection changes this distribution as well. This makes a direct comparison of $b_{\rm corr}$ between energies or event classes difficult if the aim is to isolate the emission dynamics of a source.

The need to separate intrinsic fluctuations from variations of the system size has a long history. The $\Phi$ measure proposed in Ref.~\cite{GazdzickiMrowczynski1992} was designed to preserve the single-source fluctuation signal under an independent superposition of sources. In Ref.~\cite{Gorenstein2011} this idea was developed further and the general class of strongly intensive quantities was introduced. These quantities are independent of both the mean volume and volume fluctuations in the specified statistical model, or of the mean number and number fluctuations of independent identical sources. In Ref.~\cite{Gazdzicki2013} it was explained how to introduce a common normalization with a unit independent-particle reference and a zero no-fluctuation reference.

The application of these measures to multiplicities in separated rapidity windows was proposed in Ref.~\cite{Andronov2015TMP}. For disjoint windows, the normalized second-order quantities are
\begin{align}
 \Sigma[n_F,n_B]&=
 \frac{\langle n_B\rangle\omega[n_F]+\langle n_F\rangle\omega[n_B]
    -2\operatorname{cov}(n_F,n_B)}
   {\langle n_F\rangle+\langle n_B\rangle},\label{eq:intro_sigma}\\
 \Delta[n_F,n_B]&=
 \frac{\langle n_B\rangle\omega[n_F]-\langle n_F\rangle\omega[n_B]}
   {\langle n_B\rangle-\langle n_F\rangle}.\label{eq:intro_delta}
\end{align}
Both cancel the source-number contribution present in Eq.~\eqref{eq:intro_bsources}, under the same source assumptions:
\begin{align}
 \Sigma[n_F,n_B]&=
 \Sigma[\mu_F,\mu_B],\label{eq:intro_sigma_indep}\\
 \Delta[n_F,n_B]&=
 \Delta[\mu_F,\mu_B].\label{eq:intro_delta_indep}
\end{align}

The normalization of $\Delta$ requires unequal mean multiplicities, while $\Sigma$ remains well defined for symmetric windows. Their unit independent-particle reference supplies a common baseline for comparisons across energies and collision systems.

This approach has been adopted in experimental FB analyses. NA61/SHINE studied $\Sigma[n_F,n_B]$ in $pp$ and Be+Be collisions at SPS energies \cite{ProkhorovaNA612018,AndronovNA612018}, including its dependence on the separation of the observation windows. ALICE results for  pp, Xe+Xe, and Pb+Pb collisions \cite{ErokhinIS2021,SputowskaALICE2019,SputowskaALICE2022} showed that $\Sigma$ was much less sensitive than $b_{\rm corr}$ to the width of the centrality bin and to the centrality estimator. These observations support its use for comparisons in which volume fluctuations would otherwise obscure the signal. They do not imply that $\Sigma$ must be constant with centrality: a change in the sources themselves can remain visible after the source-number contribution has been suppressed.

Preparatory model and feasibility studies have also explored these observables for future measurements at lower collision energies. Calculations for MPD at NICA and for FAIR examined strongly intensive observables within the accessible rapidity and momentum ranges, including detector response and reconstruction \cite{AltsybeevMPD2021,AliFAIR2024}.

The source-level interpretation can also be applied to strongly intensive observables. In Ref. \cite{AndronovVechernin2019} $\Sigma$ was expressed through the pair correlation function of a single string and predictions were made for $pp$ collisions at LHC energies. Charge-dependent correlations and mixtures of fused strings were investigated in Ref.~\cite{VecherninAndronov2019}. Further string-cluster and interacting string calculations examined the energy and event-selection dependencies \cite{Belokurova:2022lcr,BelokurovaVechernin2023,Prokhorova2023}. With several source types, the relative contributions can change between event classes. Cancellation that is present for a fixed identical-source law does not remove this variation. A complementary example is provided by wounded-nucleon and wounded-quark calculations \cite{Sputowska2023Theory,Sputowska2025Theory}. These studies relate $\Sigma$ to the emission profile of a wounded constituent and show that forward- and backward-moving sources need not satisfy the identical-source assumption. In these settings $\Sigma$ is not strictly strongly intensive, despite retaining useful stability properties for symmetric collisions.

Higher-order observables are of particular interest because pair correlations do not determine the full joint multiplicity distribution. Production laws with the same $\Sigma$ and $\Delta$ can have different three- and four-particle correlations. A system of all-order composition equations for moments and cumulants in superposition models was derived in Ref.~\cite{Broniowski2017}. It was suggested that solving them is a way to construct higher-order strongly intensive measures. The third-order quantity $\Gamma$ from Ref. \cite{Broniowski2017} was investigated for FB multiplicities and tested with the PYTHIA event generator \cite{Pythia83} in Ref.~\cite{Andronov2025}. A general construction of strongly intensive cumulants at arbitrary order was also presented in Ref.~\cite{Sangaline2015}. These results establish that the method need not be restricted to variances and covariances.

The question addressed here is how to retain all independent higher-order information while removing the dependence on the source-count distribution and keeping a common normalization. We construct a complete basis $J_{ij}$ with $n$ new algebraically independent observables at order $n$. The construction supplies closed-form expressions, a regular equal-mean limit, and an invertible recursion. The completeness statement applies to rational invariants of the specified composition model.

Section~\ref{sec:definitions} defines the observables and compares them with cumulant ratios. Section~\ref{sec:models} develops analytical particle production models, and Section~\ref{sec:pythia} presents the PYTHIA calculations and conditional ALICE scenarios. Section~\ref{sec:discussion} discusses the scope and limitations of the results.

Appendix~\ref{app:proof} presents the all-order construction and completeness proof. Appendix~\ref{app:conversions} collects the moment and cumulant conversion formulas. Appendix~\ref{app:vechernin} provides the evaluation prescription and symmetric-window limit for the pair-only string model. Appendix~\ref{app:production} documents the PYTHIA tune settings and auxiliary acceptance-correction procedure. Finally, Appendix~\ref{app:gaussian_explicit} gives the explicit window cumulants and evaluation formulas for the fitted Gaussian correlation hierarchy.
\section{\label{sec:definitions}Definitions and the independent-source assumption}
Let $n_F$ and $n_B$ count particles in two disjoint windows. We keep the cumulant notation of Ref.~\cite{Broniowski2017} throughout: $P_{ij}$ refers to joint cumulants of the event multiplicities, $R_{ij}$ to joint cumulants of one source, and $Q_k$ to cumulants of the source count $S$. For example, $P_{10}=\langle n_{F}\rangle$, $P_{20}=\operatorname{Var}(n_{F})$, and $P_{11}=\operatorname{cov}(n_{F},n_{B})$. The indices count the orders in $n_{F}$ and $n_{B}$.

In the independent-source model all sources share the same emission law into both windows, and this law does not depend on their number in the event. The cumulant generating functions then obey
\begin{equation}
 P(x,y)=Q\bigl(R(x,y)\bigr),\qquad
 P(x,y)=\sum_{i+j\ge1}P_{ij}\frac{x^iy^j}{i!j!}.
 \label{eq:composition}
\end{equation}
The functions $P$ and $R$ generate event and source cumulants, respectively, whereas $Q$ generates cumulants of $S$. Table~\ref{tab:notation} summarizes the notation used below. Differentiation gives the first- and second-order relations
\begin{align}
 P_{10}&=Q_1R_{10},\nonumber\\
 P_{01}&=Q_1R_{01},\nonumber\\
 P_{20}&=Q_1R_{20}+Q_2R_{10}^2,\nonumber\\
 P_{11}&=Q_1R_{11}+Q_2R_{10}R_{01},\nonumber\\
 P_{02}&=Q_1R_{02}+Q_2R_{01}^2.
 \label{eq:second}
\end{align}
At third order,
\begin{align}
 P_{30}&=Q_1R_{30}+3Q_2R_{20}R_{10}+Q_3R_{10}^3,\nonumber\\
 P_{21}&=Q_1R_{21}+Q_2(R_{20}R_{01}+2R_{11}R_{10})\nonumber\\
    &\quad+Q_3R_{10}^2R_{01},\nonumber\\
 P_{12}&=Q_1R_{12}+Q_2(R_{02}R_{10}+2R_{11}R_{01})\nonumber\\
    &\quad+Q_3R_{10}R_{01}^2,\nonumber\\
 P_{03}&=Q_1R_{03}+3Q_2R_{02}R_{01}+Q_3R_{01}^3.
\end{align}
At fourth order,
\begin{align}
 P_{40}&=Q_1R_{40}+Q_2(4R_{30}R_{10}+3R_{20}^2)\nonumber\\
    &\quad+6Q_3R_{20}R_{10}^2+Q_4R_{10}^4,\nonumber\\
 P_{31}&=Q_1R_{31}\nonumber\\
    &\quad+Q_2(R_{30}R_{01}+3R_{21}R_{10}+3R_{20}R_{11})\nonumber\\
    &\quad+3Q_3(R_{20}R_{10}R_{01}+R_{11}R_{10}^2)\nonumber\\
    &\quad+Q_4R_{10}^3R_{01},\nonumber\\
 P_{22}&=Q_1R_{22}\nonumber\\
    &\quad+Q_2(2R_{21}R_{01}+2R_{12}R_{10})\nonumber\\
    &\quad+Q_2(R_{20}R_{02}+2R_{11}^2)\nonumber\\
    &\quad+Q_3(R_{20}R_{01}^2+R_{02}R_{10}^2)\nonumber\\
    &\quad+4Q_3R_{11}R_{10}R_{01}\nonumber\\
    &\quad+Q_4R_{10}^2R_{01}^2,\nonumber\\
 P_{13}&=Q_1R_{13}\nonumber\\
    &\quad+Q_2(R_{03}R_{10}+3R_{12}R_{01}+3R_{02}R_{11})\nonumber\\
    &\quad+3Q_3(R_{02}R_{10}R_{01}+R_{11}R_{01}^2)\nonumber\\
    &\quad+Q_4R_{10}R_{01}^3,\nonumber\\
 P_{04}&=Q_1R_{04}+Q_2(4R_{03}R_{01}+3R_{02}^2)\nonumber\\
    &\quad+6Q_3R_{02}R_{01}^2+Q_4R_{01}^4.\label{eq:fourth}
\end{align}
We seek combinations independent of $Q_k$. Let us define
\begin{equation}
 m=P_{10}+P_{01},\qquad \alpha =P_{10}/m,\qquad \beta =1-\alpha .
 \label{eq:means}
\end{equation}
Thus $\alpha$ and $\beta$ are dimensionless ensemble mean fractions, with $\alpha+\beta=1$. We assume positive means in both windows, so that $m>0$ and $0<\alpha ,\beta <1$. All generating-function manipulations may be understood as operations on formal truncated series.

\begin{table*}[t]
\caption{List of notations.}
\label{tab:notation}
\begin{ruledtabular}\begin{tabular}{lp{.65\textwidth}}
Symbol & Meaning\\
$n_F,n_B$ & Event particle counts in the forward and backward windows.\\
$\mu_F,\mu_B$ & Particle counts from one source in the forward and backward windows.\\
$S$; $Q(v),Q_k$ & Source count; its cumulant generating function and $k$th cumulant.\\
$P(x,y),P_{ij}$ & Event cumulant generating function and joint cumulants.\\
$R(x,y),R_{ij}$ & Single-source cumulant generating function and joint cumulants.\\
$m$; $\alpha,\beta$ & Total mean count; forward and backward mean fractions, Eq.~\eqref{eq:means}.\\
$D(v)$ & Diagonal series $P(v,v)$ generating total-multiplicity cumulants.\\
$h(t,u)$ & Invariant series defined by Eq.~\eqref{eq:h_main}.\\
$J_{ij}$ & Normalized strongly intensive coordinates, Eq.~\eqref{eq:algorithm_output}.\\
$X$; $X_{\rm s}$ & Mean-balanced event and source differences, Eq.~\eqref{eq:contrast_cumulants}.\\
$f_n(X)$ & Formal scalar factorial cumulants of the signed variable $X$.\\
$F_n$ & Weighted particle factorial cumulants, Eq.~\eqref{eq:weighted_factorial}.\\
$f_{ij}$ & Joint factorial cumulants of $(n_F,n_B)$.\\
$p_n(z)$ & Negative-binomial cumulant polynomials, Eq.~\eqref{eq:nb_polynomials}.\\
$c_2,c_4$ & Polynomial density-profile coefficients.\\
$a_2,a_3,a_4$ & Connected-correlation amplitudes of the Gaussian hierarchy.\\
$H_{ij}$ & Window integrals of connected factorial densities.\\
$\mathcal K_{ij}$ & Normalized directional source cumulants, Eq.~\eqref{eq:gauss_C_def}.\\
$r_3,r_4$ & Amplitude ratios used in the transfer prescription, Sec.~\ref{sec:pythia_alice}.\\
\end{tabular}\end{ruledtabular}
\end{table*}

\subsection{\label{sec:subsecIIA}The known strongly intensive observables}
In this notation, the conventional second- and third-order strongly intensive observables are
\begin{align}
 \Sigma&=\frac{\beta ^2P_{20}+\alpha ^2P_{02}-2\alpha \beta P_{11}}{m\alpha \beta },
 \label{eq:sigma}\\
 \Delta&=\frac{\beta ^2P_{20}-\alpha ^2P_{02}}{m\alpha \beta (\beta -\alpha )},
 \label{eq:delta}\\
 \Gamma&=\frac{\beta ^3P_{30}-3\alpha \beta ^2P_{21}+3\alpha ^2\beta P_{12}-\alpha ^3P_{03}}
 {m\alpha \beta (\beta -\alpha )}.
 \label{eq:gamma}
\end{align}
These definitions agree with Refs.~\cite{Gazdzicki2013,Andronov2015TMP,Andronov2025}. The independent-particle reference used here assigns each particle in the union of the windows independently to the forward or backward window, with probabilities $\alpha $ and $\beta $, so that the reference means coincide with the event means. The union multiplicity may have any distribution. All three observables equal one in this reference.

The $\Delta$ and $\Gamma$ normalizations require unequal mean multiplicities ($\beta \neq \alpha $). If the joint distribution is invariant under $F\leftrightarrow B$, then $\Sigma$ reduces to
\begin{equation}
\Sigma=\frac{P_{20}-P_{11}}{P_{10}}.
\label{eq:sigma_sym}
\end{equation}

\subsection{\label{sec:subsecIIB}The fourth-order observable}
A direct fourth-order extension of the zero-mean contrast $X_s=\beta\mu_F-\alpha\mu_B$ used in $\Gamma$ is not strongly intensive. Its fourth cumulant contains $3Q_2\kappa_2(\beta \mu_F-\alpha \mu_B)^2$, which survives division by the mean multiplicity. Removing this contribution requires a term quadratic in second-order cumulants. One such normalized combination is

\begin{align}
\Upsilon=\frac{1}{3m}\bigg\{&m^2\Sigma^2
 \left(\frac{P_{20}}{P_{10}^2}+\frac{P_{11}}{P_{10}P_{01}}
             +\frac{P_{02}}{P_{01}^2}\right)\nonumber\\
 &-\frac{P_{40}P_{01}^2}{P_{10}^2}
 -\frac{P_{04}P_{10}^2}{P_{01}^2}
 +4\frac{P_{31}P_{01}}{P_{10}}\nonumber\\
 &+4\frac{P_{13}P_{10}}{P_{01}}-6P_{22}\bigg\}.
\label{eq:upsilon}
\end{align}

$\Upsilon$ equals one if particles enter the two windows independently, whatever the distribution of the union multiplicity, and zero if both window multiplicities are fixed event by event. If the joint distribution is invariant under $F\leftrightarrow B$, so that $P_{ij}=P_{ji}$, $\Upsilon$ reduces to:
\begin{align}
\Upsilon&=\frac{{\left(P_{20}-P_{11}\right)}^{2}\left(4P_{20}+2P_{11}\right)}{3P_{10}^{3}}\nonumber\\
 &-\frac{P_{40}-4P_{31}+3P_{22}}{3P_{10}}.
\label{eq:upsilon_sym}
\end{align}

\subsection{\label{sec:subsecIIC}The complete basis}
The composition equations admit additional invariants beyond $\Sigma$, $\Delta$, $\Gamma$, and $\Upsilon$. Arbitrary-order constructions were developed in Refs.~\cite{Broniowski2017,Sangaline2015}. Here we require the new observables to have the reference values of Sec.~\ref{sec:subsecIIA}---unity for independent particles and zero for fixed multiplicities---and no singularities at equal positive means.

The construction uses two functions: the diagonal restriction $D(v)=P(v,v)$, which generates the cumulants of the union multiplicity $n_F+n_B$, and the series
\begin{equation}
 h(t,u)=D^{-1}\!\left(P(\beta t+u,-\alpha t+u)\right),
 \label{eq:h_main}
\end{equation}
where the weights are fixed to the ensemble fractions $\alpha=P_{10}/m$ and $\beta=P_{01}/m$. The linear part of the argument is $mu$. It carries the \emph{event multiplicity asymmetry}
\begin{equation}
 X=\beta n_F-\alpha n_B ,
 \label{eq:asymmetry}
\end{equation}
whose first cumulant vanishes by construction:
\begin{equation}
 \langle X\rangle=\beta\,\alpha m-\alpha\,\beta m=0 ,
 \label{eq:asymmetry_mean}
\end{equation}
Since $D'(0)=m\ne0$, the formal inverse exists. Under $P=Q\circ R$, the outer function $Q$ enters both $D$ and the argument of Eq.~\eqref{eq:h_main} and cancels there. Normalized derivatives of $e^h$, supplemented by multiples of $\Sigma$, define $J_{ij}$ as specified below in Eq.~\eqref{eq:algorithm_output}. Appendix~\ref{app:proof} gives the coefficient recursion and its inverse.

Through fourth order the complete list is
\begin{equation}
\begin{array}{c|l}
 \text{order}&\text{new observables}\\ \hline
 2&J_{20}=\Sigma,\quad J_{11}\\
 3&J_{30},\quad J_{21},\quad J_{12}\\
 4&\Upsilon,\quad J_{31},\quad J_{22},\quad J_{13}.
\end{array}
 \label{eq:list}
\end{equation}
The explicit expressions through third order are
\begin{align}
 J_{20}={}&\Sigma,\\
 J_{11}={}&\Sigma+
 \frac{\beta (P_{20}+P_{11})-\alpha (P_{11}+P_{02})}{m\alpha \beta },\\
 J_{30}={}&\frac{\beta ^3P_{30}-3\alpha \beta ^2P_{21}+3\alpha ^2\beta P_{12}-\alpha ^3P_{03}}{m\alpha \beta }
 +2\alpha \Sigma,\\
 J_{21}={}&\frac{\beta ^2P_{30}+(\beta ^2-2\alpha \beta )P_{21}
 +(\alpha ^2-2\alpha \beta )P_{12}+\alpha ^2P_{03}}{m\alpha \beta }\nonumber\\
 &+\left(1-\frac{P_{20}+2P_{11}+P_{02}}m\right)\Sigma,\\
 J_{12}={}&\Sigma+
 \frac{\beta P_{30}+(2\beta -\alpha )P_{21}+(\beta -2\alpha )P_{12}-\alpha P_{03}}{m\alpha \beta }
 \nonumber\\
 &+2\left(1-\frac{P_{20}+2P_{11}+P_{02}}m\right)(J_{11}-\Sigma).
\end{align}
Every member is dimensionless, equals one for independent particles, and vanishes when both multiplicities are fixed. All remain defined at equal positive means. When $\alpha \ne \beta $, two of them can be written as
\begin{align}
 J_{11}&=\Sigma+\frac{\beta -\alpha }{2\alpha \beta }(\Delta-\Sigma),\label{eq:j11}\\
 J_{30}&=(\beta -\alpha )\Gamma+2\alpha \Sigma.\label{eq:j30}
\end{align}

We use $\Upsilon$ in place of $J_{40}$ in the numerical results. The canonical fourth-order coordinate is
\begin{align}
 J_{40}={}&\frac{\beta ^4P_{40}-4\alpha \beta ^3P_{31}+6\alpha ^2\beta ^2P_{22}
 -4\alpha ^3\beta P_{13}+\alpha ^4P_{04}}{m\alpha \beta }\nonumber\\
 &+3\alpha \beta \left(1-\frac{P_{20}+2P_{11}+P_{02}}m\right)\Sigma^2
 +3\alpha \beta \Sigma.
\end{align}
The replacement preserves completeness because
\begin{align}
 J_{40}={}&-3\alpha \beta \Upsilon+3\alpha \beta \Sigma^2+3\alpha \beta \Sigma\nonumber\\
 &+3\alpha \beta (\beta -\alpha )(J_{11}-\Sigma)\Sigma^2
 +(1-3\alpha \beta )\Sigma^3.
 \label{eq:upsilon_conversion}
\end{align}
The coefficient of $\Upsilon$ is nonzero for $\alpha \beta \ne0$. The remaining fourth-order coordinates are
\begin{widetext}
\begin{align}
 J_{31}={}&\frac{\beta ^3P_{40}+(\beta ^3-3\alpha \beta ^2)P_{31}
 +(3\alpha ^2\beta -3\alpha \beta ^2)P_{22}+(3\alpha ^2\beta -\alpha ^3)P_{13}-\alpha ^3P_{04}}{m\alpha \beta }
 \nonumber\\
 &+\left(1-\frac{P_{20}+2P_{11}+P_{02}}m\right)
 \left[J_{30}-2\alpha \Sigma+3\alpha \beta \Sigma(J_{11}-\Sigma)\right]+2\alpha \Sigma,
 \\
 J_{22}={}&\frac{\beta ^2P_{40}+(2\beta ^2-2\alpha \beta )P_{31}
 +(\beta ^2+\alpha ^2-4\alpha \beta )P_{22}+(2\alpha ^2-2\alpha \beta )P_{13}+\alpha ^2P_{04}}{m\alpha \beta }
 \nonumber\\
 &+2\left(1-\frac{P_{20}+2P_{11}+P_{02}}m\right)
 \left[J_{21}+\alpha \beta (J_{11}-\Sigma)^2\right]
 \nonumber\\
 &+\left[-1+2\frac{P_{20}+2P_{11}+P_{02}}m
 -\frac{P_{30}+3P_{21}+3P_{12}+P_{03}}m\right]\Sigma,
 \\
 J_{13}={}&\Sigma+
 \frac{\beta P_{40}+(3\beta -\alpha )P_{31}+3(\beta -\alpha )P_{22}+(\beta -3\alpha )P_{13}-\alpha P_{04}}{m\alpha \beta }
 \nonumber\\
 &+3\left(1-\frac{P_{20}+2P_{11}+P_{02}}m\right)(J_{12}-\Sigma)
 \nonumber\\
 &+3\left[2\frac{P_{20}+2P_{11}+P_{02}}m
 -\frac{P_{30}+3P_{21}+3P_{12}+P_{03}}m-1\right](J_{11}-\Sigma).
\end{align}
\end{widetext}

Appendix~\ref{app:proof} gives a constructive proof of the following theorem:
\begin{theorem}
Through order $n$, the mean fraction $\alpha$ and the quantities $J_{ij}$ with $i\ge1$, $j\ge0$, and $2\le i+j\le n$ form a complete algebraically independent set of rational invariants of the invertible outer composition of Eq.~\eqref{eq:composition}. For $n\ge2$, the hierarchy contains $n$ observables at order $n$ and $n(n+1)/2-1$ in total.
\end{theorem}
All $J_{ij}$ are constructed in this work with the normalization of Sec.~\ref{sec:subsecIIA}: unity for independent particles, zero for fixed multiplicities, and regularity at equal positive means. At second order, $J_{20}=\Sigma$ and $J_{11}$ is a linear combination of $\Sigma$ and $\Delta$, Eq.~\eqref{eq:j11}. From third order on, the observables involve cumulants of new orders. The theorem concerns formal cumulant series and holds on the domain $m\alpha\beta\ne0$. In particular, the domain includes equal positive means. Completeness refers to invariants of formal outer composition and does not imply that every assignment of invariant values arises from a physical multiplicity distribution.

Operationally, changing the source-count distribution at fixed source emission law leaves $\alpha$ and all $J_{ij}$ unchanged. Conversely, at a fixed truncation order, two formal cumulant series with the same $\alpha$ and all $J_{ij}$ through that order differ only by an invertible outer composition. Thus these coordinates retain all rationally invariant cumulant information through the measured order, without reconstructing a unique source emission law or establishing that the connecting outer series is a physically admissible source-count generating function.

\subsection{Comparison with asymmetry cumulants}
Higher-order fluctuation observables are particularly useful in the search for the quantum chromodynamics critical point: their critical contributions can have a stronger dependence on the correlation length than the variance \cite{Stephanov2009Critical,Athanasiou:2010kw}. Experiments therefore measure cumulants of net electric charge and net-proton number, together with factorial cumulants (integrated connected particle correlations) of the underlying multiplicities \cite{STARNetCharge2014,STARProtonCorrelations2021}. Net charge is constructed as $N_+-N_-$, where $N_+$ and $N_-$ count positively and negatively charged particles in the same acceptance. Similarly, net-proton number is $N_p-N_{\bar p }$. The latter is a measured proxy for net-baryon fluctuations, not the full net-baryon number. In both cases the observable is built from two extensive particle counts.

The same construction provides a mathematical analogy with the forward-backward multiplicities $n_F$ and $n_B$: the event multiplicity asymmetry $X=\beta n_F-\alpha n_B$ of Sec.~\ref{sec:subsecIIC} is the mean-balanced analogue of the net-charge and net-proton differences, although here the two labels are spatial windows rather than charge or particle--antiparticle labels. It is a weighted difference, not the event-wise ratio $(n_F-n_B)/(n_F+n_B)$. Its ordinary and factorial cumulants can be computed, but they are not generally strongly intensive: they retain dependence on the source-count law $Q$. As shown below, the zero mean of $X$ removes several terms, yet its fourth ordinary cumulant still contains $Q_2$ even after division by the mean multiplicity $m$. For factorial cumulants we distinguish between two different objects: a formal factorial cumulant of the signed variable $X$, and a weighted combination of the factorial cumulants of the particle numbers themselves.

We now take cumulants of $X$, treating $\alpha$ and $\beta$ as fixed constants. One obtains
\begin{align}
 \kappa_1(X)&=\beta P_{10}-\alpha P_{01}=0,\nonumber\\
 \kappa_2(X)&=\beta ^2P_{20}-2\alpha \beta P_{11}+\alpha ^2P_{02},\nonumber\\
 \kappa_3(X)&=\beta ^3P_{30}-3\alpha \beta ^2P_{21}+3\alpha ^2\beta P_{12}-\alpha ^3P_{03},\nonumber\\
 \kappa_4(X)&=\beta ^4P_{40}-4\alpha \beta ^3P_{31}+6\alpha ^2\beta ^2P_{22}\nonumber\\
 &\quad-4\alpha ^3\beta P_{13}+\alpha ^4P_{04}.
 \label{eq:contrast_cumulants}
\end{align}
In particular $\kappa_2(X)/m=\alpha \beta \Sigma$ and $\kappa_3(X)/m=\alpha \beta (J_{30}-2\alpha \Sigma)$.

For the source contrast $X_s=\beta\mu_F-\alpha\mu_B$ of Sec.~\ref{sec:subsecIIB}, the random-sum formulas simplify because $\langle X_s\rangle=0$, and $m=Q_1 m_s$, where $m_s$ is the mean multiplicity of one source:
\begin{align}
 \kappa_1(X)&=0,\nonumber\\
 \kappa_2(X)&=Q_1\kappa_2(X_{\rm s}),\nonumber\\
 \kappa_3(X)&=Q_1\kappa_3(X_{\rm s}),\nonumber\\
 \kappa_4(X)&=Q_1\kappa_4(X_{\rm s})+3Q_2\kappa_2(X_{\rm s})^2.
\end{align}
Division by $m$ therefore removes the dependence on $Q_1$ through third order in this zero-mean direction, but fails at fourth order.

For a signed, noninteger weighted quantity such as $X$, the following two conventions must not be confused. The scalar falling-factorial cumulants are defined formally by derivatives of $\log\langle(1+z)^X\rangle$, understood as a formal power series in $z$. We denote them by $f_n(X)$. Through fourth order,
\begin{align}
 f_1(X)&=\kappa_1(X)=0,\nonumber\\
 f_2(X)&=\kappa_2(X)-\kappa_1(X)=\kappa_2(X),\nonumber\\
 f_3(X)&=\kappa_3(X)-3\kappa_2(X)+2\kappa_1(X)\nonumber\\
    &=\kappa_3(X)-3\kappa_2(X),\nonumber\\
 f_4(X)&=\kappa_4(X)-6\kappa_3(X)+11\kappa_2(X)-6\kappa_1(X)\nonumber\\
    &=\kappa_4(X)-6\kappa_3(X)+11\kappa_2(X).
\end{align}
These quantities are mathematically well defined but do not count connected correlations of distinct particles carrying the weights $\beta$ and $-\alpha$. For that purpose, one uses instead the derivatives of $\log\langle(1+\beta z)^{n_F}(1-\alpha z)^{n_B}\rangle$ \cite{KitazawaLuo2017Factorial}. We denote them by $F_n$ and the joint factorial cumulants of $(n_F,n_B)$ by $f_{ij}$. Then
\begin{align}
 F_1&=\beta f_{10}-\alpha f_{01}=0,\nonumber\\
 F_2&=\beta ^2f_{20}-2\alpha \beta f_{11}+\alpha ^2f_{02},\nonumber\\
 F_3&=\beta ^3f_{30}-3\alpha \beta ^2f_{21}+3\alpha ^2\beta f_{12}-\alpha ^3f_{03},\nonumber\\
 F_4&=\beta ^4f_{40}-4\alpha \beta ^3f_{31}+6\alpha ^2\beta ^2f_{22}\nonumber\\
  &\quad-4\alpha ^3\beta f_{13}+\alpha ^4f_{04}.
 \label{eq:weighted_factorial}
\end{align}
The same structure holds because $F_1(X_s)=0$. The general conversion from $P_{ij}$ to $f_{ij}$ is given in Appendix~\ref{app:conversions}. A useful check is $F_2=\kappa_2(X)-m\alpha \beta =m\alpha \beta (\Sigma-1)$. Independent Poisson $n_F,n_B$ have $F_n=0$ for $n\ge2$, whereas their scalar $f_n(X)$ need not vanish. Thus these definitions are not interchangeable, even at second order.

For an unweighted multiplicity or a net multiplicity, commonly used ordinary ratios (when the denominators are nonzero) include $\kappa_2/\kappa_1$, $\kappa_3/\kappa_2$, and $\kappa_4/\kappa_2$~\cite{STARProtonCorrelations2021}. For the proton count $N_p$ itself, factorial ratios such as $f_2(N_p)/f_1(N_p)$ and $f_3(N_p)/f_1(N_p)$ are also measured~\cite{STARPrecision2025}. $f_4(N_p)/f_1(N_p)$ provides the analogous fourth-order normalization. Here $f_1(N_p)=\langle N_p\rangle>0$: the argument is the proton count, not the weighted contrast $X$. At a fixed number of identical independent sources, all these cumulants are proportional to that number, so ratios remove it. This statement does not generally survive fluctuations of that number.

Our event multiplicity asymmetry has zero mean by construction. Ratios with $\kappa_1$, $f_1$, or $F_1$ in the denominator are therefore undefined. When their denominators are nonzero, $\kappa_3/\kappa_2$ and $F_3/F_2$ are source-count independent in this zero-mean model. However,
\begin{align}
 &\frac{\kappa_4(X)}{\kappa_2(X)}
 =\frac{\kappa_4(X_{\rm s})}{\kappa_2(X_{\rm s})}
    +3\frac{Q_2}{Q_1}\kappa_2(X_{\rm s}),\\
 &\frac{F_4}{F_2}
 =\frac{F_{4,\mathrm{source}}}{F_{2,\mathrm{source}}}
    +3\frac{Q_2}{Q_1}F_{2,\mathrm{source}}.
\end{align}

Table~\ref{tab:ratios} summarizes the distinction between the normalized invariants and common ratios in the zero-mean direction.
\begin{table*}[t]
\caption{Dependence on source-count fluctuations for $X=\beta n_F-\alpha n_B$ and the corresponding weighted factorial cumulants $F_n$. The independent identical-source assumptions apply. Ratios require nonzero denominators.}
\label{tab:ratios}
\begin{ruledtabular}\begin{tabular}{ll}
Quantity & Dependence on the source-count law\\
$\kappa_3(X)/\kappa_2(X)$ & None.\\
$F_3/F_2$ & None.\\
$\kappa_4(X)/\kappa_2(X)$ & Contains $3(Q_2/Q_1)\kappa_2(X_{\rm s})$.\\
$F_4/F_2$ & Contains $3(Q_2/Q_1)F_{2,\mathrm{source}}$.\\
$J_{ij}$ & None at every order.\\
\end{tabular}\end{ruledtabular}
\end{table*}

Several procedures aim to reduce this dependence on $Q$. A widely used example is the centrality bin width correction (CBWC) \cite{Luo:2013bmi}: cumulants are evaluated in narrow bins of a reference multiplicity and then averaged with event-count weights. Fixing the reference multiplicity, however, does not in general fix the number of sources. Residual size fluctuations may remain within each bin. Moreover, conditioning on a reference multiplicity correlated with the measured particles can modify the dynamical fluctuations themselves. CBWC is not a general algebraic cancellation of $Q$, and its validity must be established for the relevant production and selection conditions. Both successful and misleading CBWC results in toy models were demonstrated in Ref.~\cite{FrimanKoch2026CBWC}.

The same algebra can be applied to two species counts, such as proton and antiproton multiplicities, provided the independent identical-source assumptions hold. It does not by itself identify the resulting invariants with net-baryon or net-charge susceptibilities. Global conservation and acceptance effects require separate treatment: reduced factorial cumulants provide complementary acceptance tests \cite{Bzdak2025}, and the relation between coordinate-space and momentum-space fluctuations depends on the dynamics \cite{Kuznietsov2024Cuts}. When the antiproton mean vanishes, the present normalization is undefined. A small nonzero mean can also make estimation unstable because the normalizations divide by powers of the window means. A pair of more abundant species can instead be used to study higher-order chemical fluctuations \cite{NA49:2011ffk,STAR:2014nuj,ALICE:2017jsh,NA61SHINE:2020cxu} where the same set of $J_{ij}$ provides an alternative set of fluctuation measures.
\section{\label{sec:models}Analytical models}
We now evaluate the hierarchy in analytically tractable emission models. For sources with a common single-source law $R$, it suffices to apply the construction at the level of $R$: by the composition invariance of Sec.~\ref{sec:subsecIIC}, the source-count law $Q$ cancels from all $J_{ij}$ and from $\Sigma$, $\Delta$, $\Gamma$, and $\Upsilon$.

Consider a random number $S$ of identical, independent sources. Each source emits into the two disjoint windows with mean multiplicities $a=R_{10}=\langle\mu_F\rangle>0$ and $b=R_{01}=\langle\mu_B\rangle>0$. The disjoint-window and independent-source assumptions of Sec.~\ref{sec:definitions} apply. The windows may have different sizes, so $a$ and $b$ are kept arbitrary. The fractions $\alpha=a/(a+b)$ and $\beta=b/(a+b)$ coincide with the ensemble fractions of Sec.~\ref{sec:definitions}, since $P_{10}=Q_1a$ and $m=Q_1(a+b)$. The distribution of $S$ is arbitrary, with positive mean and finite cumulants through the order considered. It may be fixed, Poisson, negative binomial, or another count law.
\subsection{Independent Poisson emission into two windows}
\label{sec:independent_emission}
First suppose that the two multiplicities of a single source are independent Poisson variables. Their cumulant generating function is
\begin{equation}
 R(x,y)=a(e^x-1)+b(e^y-1).
 \label{eq:independent_poisson_R}
\end{equation}
Thus $R_{i0}=a$ and $R_{0j}=b$ for $i,j\ge1$, and all mixed source cumulants vanish. The event multiplicities are generally correlated: for example, $P_{11}=ab\,\mathrm{Var}(S)$. The $J_{ij}$ construction removes the entire source-count law. For this source its defining exponential is simply $e^u(\alpha e^{\beta t}+\beta e^{-\alpha t})$. Consequently,
\begin{align}
 &J_{ij}=1 &&(i\ge1,\ i+j\ge2),\nonumber\\
 &\Sigma=\Delta=\Gamma=\Upsilon=1.
 \label{eq:independent_poisson_answer}
\end{align}
The statement for $J_{ij}$ holds at every order. The equalities for $\Delta$ and $\Gamma$ apply only when $\alpha\ne\beta$.
\subsection{Independent negative-binomial emission}
\label{sec:independent_nbd}
The Poisson model returns unity at every order, so any departure of the observables from one must originate in the emission law itself. The negative binomial is one of the simplest such laws and a standard parametrization of multiplicity fluctuations. Keep the two multiplicities of a source independent, but replace each Poisson variable by an independent negative-binomial variable with means $a,b$ and positive shape parameters $k_A,k_B$:
\begin{align}
 R(x,y)={}&-k_A\log[1-(a/k_A)(e^x-1)]\nonumber\\
     &-k_B\log[1-(b/k_B)(e^y-1)].
 \label{eq:independent_nbd_R}
\end{align}
There is no shared fluctuating intensity between the two windows at the single-source level in this model. In particular, all mixed $R_{ij}$ with $i,j>0$ vanish.

To give compact explicit answers, we write $z_A=a/k_A$, $z_B=b/k_B$ and introduce three polynomials:
\begin{align}
 p_2(z)&=1+z,\nonumber\\
 p_3(z)&=1+3z+2z^2,\nonumber\\
 p_4(z)&=1+7z+12z^2+6z^3.
 \label{eq:nb_polynomials}
\end{align}
We abbreviate $p_{nA}=p_n(z_A)$, $p_{nB}=p_n(z_B)$ and $w_n=\alpha p_{nA}+\beta p_{nB}$ for $n=2,3$. They give $R_{n0}=a p_{nA}$ and $R_{0n}=b p_{nB}$. $w_n$ is the $n$th cumulant of the total source multiplicity divided by its mean. Thus the expressions below contain only $a,b,k_A,k_B$.

The conventional quantities and $\Upsilon$ are
\begin{align}
 \Sigma&=\beta p_{2A}+\alpha p_{2B}=1+\beta z_A+\alpha z_B,\nonumber\\
 \Delta&=\frac{\beta p_{2A}-\alpha p_{2B}}{\beta -\alpha },\nonumber\\
 \Gamma&=\frac{\beta ^2p_{3A}-\alpha ^2p_{3B}}{\beta -\alpha },\nonumber\\
 \Upsilon&=\frac{\Sigma^3-\beta ^3p_{4A}-\alpha ^3p_{4B}}{3\alpha \beta }.
 \label{eq:independent_nbd_named}
\end{align}
All remaining coordinates through fourth order are given explicitly by
\begin{align}
 J_{11}={}&\Sigma+p_{2A}-p_{2B},\nonumber\\
 J_{30}={}&\beta ^2p_{3A}-\alpha ^2p_{3B}+2\alpha \Sigma,\nonumber\\
 J_{21}={}&\beta p_{3A}+\alpha p_{3B}+(1-w_2)\Sigma,\nonumber\\
 J_{12}={}&\Sigma+p_{3A}-p_{3B}
    +2(1-w_2)(p_{2A}-p_{2B}),\nonumber\\
 J_{40}={}&\beta ^3p_{4A}+\alpha ^3p_{4B}
    +3\alpha \beta (1-w_2)\Sigma^2+3\alpha \beta \Sigma,\nonumber\\
 J_{31}={}&\beta ^2p_{4A}-\alpha ^2p_{4B}+2\alpha \Sigma\nonumber\\
 &+(1-w_2)[\beta ^2p_{3A}-\alpha ^2p_{3B}
       +3\alpha \beta \Sigma(p_{2A}-p_{2B})],\nonumber\\
 J_{22}={}&\beta p_{4A}+\alpha p_{4B}
    +2(1-w_2)[J_{21}+\alpha \beta (p_{2A}-p_{2B})^2]\nonumber\\
 &+(-1+2w_2-w_3)\Sigma,\nonumber\\
 J_{13}={}&\Sigma+p_{4A}-p_{4B}
    +3(1-w_2)(J_{12}-\Sigma)\nonumber\\
 &+3(2w_2-w_3-1)(p_{2A}-p_{2B}).
 \label{eq:independent_nbd_basis}
\end{align}
$J_{40}$ and $\Upsilon$ are related by Eq.~\eqref{eq:upsilon_conversion} and are alternative coordinates. These formulas show that zero inter-window correlation within a source does not imply unit strongly intensive observables: correlations among particles in the same window contribute.

A useful width-dependent specialization assumes a homogeneous source with $a=\rho\,\delta\eta_A$, $b=\rho\,\delta\eta_B$ and shape parameters $k_A=\kappa\,\delta\eta_A$, $k_B=\kappa\,\delta\eta_B$. This proportionality is an extra assumption: independent negative-binomial emission does not prescribe how the shape parameters change with the window width. Under it $z_A=z_B=z=\rho/\kappa$, while $\alpha=\delta\eta_A/(\delta\eta_A+\delta\eta_B)$ still records the relative widths. For this specialization,
\begin{equation}
 \Sigma=\Delta=J_{11}=J_{12}=J_{13}=1+z.
\end{equation}
\begin{align}
 \Gamma&=1+3z+2z^2,\nonumber\\
 J_{21}&=(1+z)^2,\nonumber\\
 J_{30}&=(\beta -\alpha )(1+3z+2z^2)+2\alpha (1+z),\nonumber\\
 J_{31}&=(\beta -\alpha )(1+6z+9z^2+4z^3)\nonumber\\
 &\quad+2\alpha (1+z),\nonumber\\
 J_{22}&=(1+z)^2(1+2z),\nonumber\\
 \Upsilon&=\frac{(1+z)^3}{3\alpha \beta }\nonumber\\
 &\quad-\frac{1-3\alpha \beta }{3\alpha \beta }(1+7z+12z^2+6z^3).
 \label{eq:independent_nbd_equal_z}
\end{align}

Within the same specialization, the equal-mean limit ($\alpha=\beta=\frac{1}{2}$) is the case where the conventional $\Delta$ and $\Gamma$ are undefined. The full hierarchy remains finite and reduces to
\begin{equation}
 \Sigma=J_{11}=J_{12}=J_{13}=J_{30}=J_{31}=1+z.
\end{equation}
\begin{align}
 &J_{21}=\Sigma^2,\nonumber\\
 &J_{22}=\Sigma^2(2\Sigma-1),\nonumber\\
 &\Upsilon=\frac{-2\Sigma^3+6\Sigma^2-\Sigma}{3}.
 \label{eq:independent_nbd_equal_z_sym}
\end{align}

Neither the Poisson nor the negative-binomial model depends on the window separation at fixed widths and fixed source parameters. Such a dependence requires inter-window correlations or a position-dependent source law, as in the string model below. The contrast between the two models is the point of this subsection: Poisson emission gives unity at all orders, whereas merely non-Poisson independent emission already moves every observable away from unity through within-window correlations.

\subsection{Independent strings with the two-particle exponential pair function}
\label{sec:independent_with_pair_corr}
The two preceding models emit into the windows independently and therefore cannot depend on the window separation at fixed widths. The minimal extension that produces such a dependence keeps the source translationally invariant but gives its correlations a finite range. Following the single-string formulation of Ref.~\cite{AndronovVechernin2019}, let the one-particle density be constant, $\lambda_1(\eta)=\mu_0$, and choose the two-particle correlation function for a string as $\Lambda\left(\eta_1-\eta_2\right)=\Lambda_0\exp(-|\eta_1-\eta_2|/\ell)$. The normalized pair function is related to the inclusive two-particle density by
\begin{equation}
  \Lambda\left(\eta_1-\eta_2\right)=\frac{\lambda_2(\eta_{1},\eta_{2})}{\lambda_1(\eta_{1})\lambda_1(\eta_{2})}-1,
\end{equation}
For a constant one-particle density,
\begin{equation}
  \Lambda\left(\eta_1-\eta_2\right)=\frac{\lambda_2(\eta_{1},\eta_{2})}{\mu_{0}^{2}}-1.
\end{equation}

Higher-order cumulants additionally require the corresponding multiparticle densities. The standard moment--cumulant expansion sums over partitions of the particles into disjoint groups \cite{DeWolf1996Correlations}. Explicitly,
\begin{align}
 \frac{\lambda_3}{\mu_0^3}
 &=1+\Lambda_{12}+\Lambda_{13}+\Lambda_{23}
  +\frac{C_3(1,2,3)}{\mu_0^3},\nonumber\\
 \frac{\lambda_4}{\mu_0^4}
 &=1+\sum_{i<j}\Lambda_{ij}
  +\Lambda_{12}\Lambda_{34}+\Lambda_{13}\Lambda_{24}\nonumber\\
 &\quad+\Lambda_{14}\Lambda_{23}
  +\frac{\sum_{i<j<k}C_3(i,j,k)}{\mu_0^3}
  +\frac{C_4(1,2,3,4)}{\mu_0^4}.
 \label{eq:general_density_partitions}
\end{align}
Here $\lambda_n$ denotes the inclusive density of $n$ distinct particles from one source, $\Lambda_{ij}=\Lambda\left(\eta_{i}-\eta_{j}\right)$. The functions $C_3$ and $C_4$ are connected factorial densities.

The source cumulants follow from integrals of the inclusive densities. For disjoint windows $F$ and $B$, define
\begin{equation}
 L_{ij}=\int_{F^i\times B^j}
 \lambda_{i+j}(\eta_1,\ldots,\eta_{i+j})
 \,d\eta_1\cdots d\eta_{i+j}.
 \label{eq:inclusive_window_integrals}
\end{equation}
The conversion from the integrated densities $L_{ij}$ to ordinary source cumulants is given in Appendix~\ref{app:conversions}.

\begin{figure*}[!t]
\centering
\includegraphics[width=\textwidth]{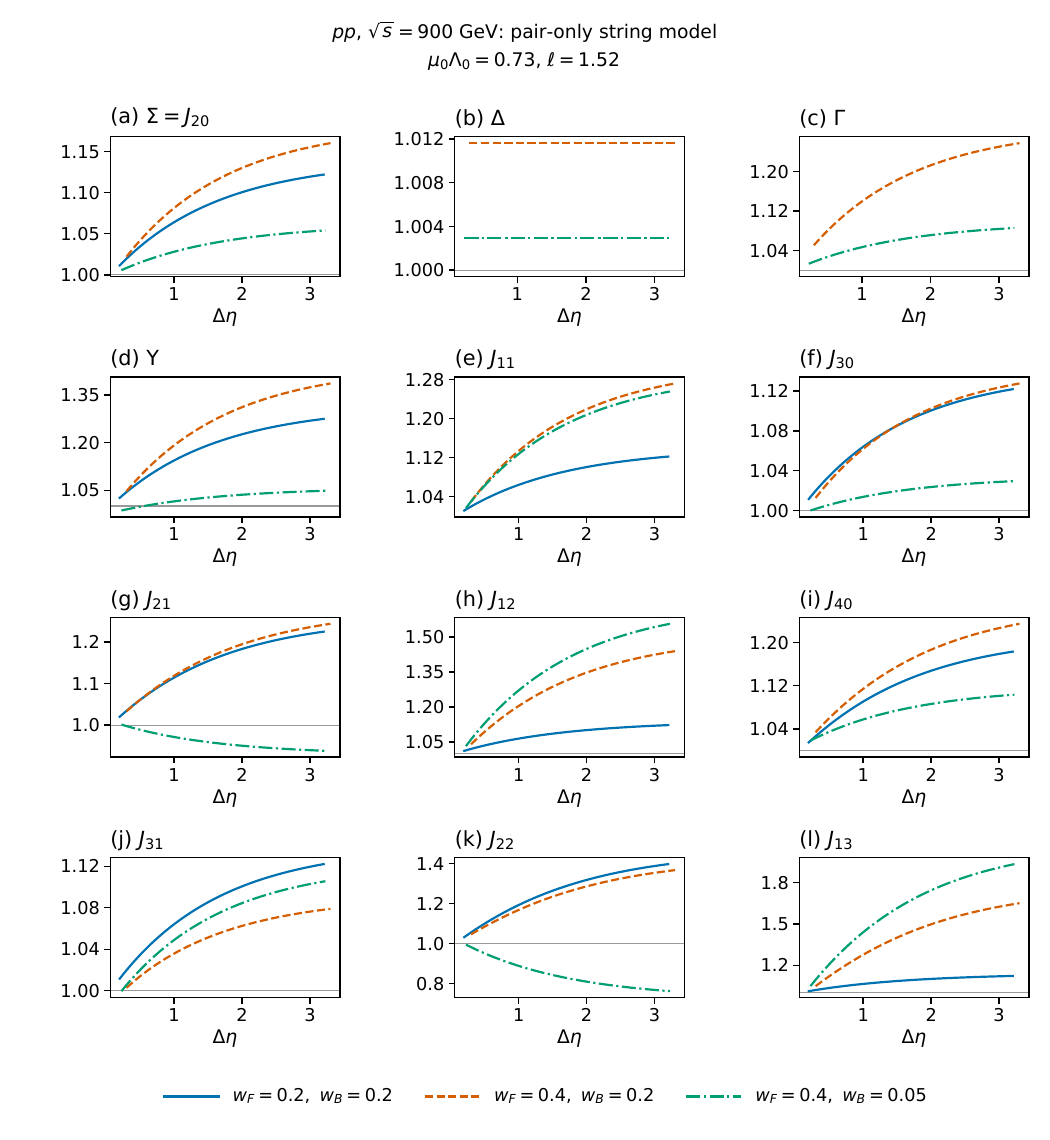}
\caption{Strongly intensive observables through fourth order as a function of the distance between window centers. Curves are obtained directly from the ordinary source cumulants in Eq.~\eqref{eq:pair_ordinary_all}, with the exponential pair parameters $\mu_0\Lambda_0=0.73$, $\ell=1.52$. Solid blue curves have widths $0.2,0.2$. Dashed orange curves have widths $0.4,0.2$. Dash-dotted green curves have widths $0.4,0.05$. Panels (a)--(l) show $\Sigma=J_{20}$, $\Delta$, $\Gamma$, $\Upsilon$, $J_{11}$, $J_{30}$, $J_{21}$, $J_{12}$, $J_{40}$, $J_{31}$, $J_{22}$, and $J_{13}$, respectively.}
\label{fig:pair_av2019}
\end{figure*}

As a pair-only closure through fourth order, set $C_3=C_4=0$, while retaining $C_2=\mu_0^2\Lambda$. Then
\begin{align}
 \frac{\lambda_3}{\mu_0^3}&=1+\Lambda_{12}+\Lambda_{13}+\Lambda_{23},\nonumber\\
 \frac{\lambda_4}{\mu_0^4}&=1+\sum_{i<j}\Lambda_{ij}
 +\Lambda_{12}\Lambda_{34}+\Lambda_{13}\Lambda_{24}
 +\Lambda_{14}\Lambda_{23}.
 \label{eq:pair_only_densities}
\end{align}
Vanishing connected factorial densities above second order do not imply vanishing higher-order $J_{ij}$: these observables also contain lower-order correlations. For the exponential pair function, define
\begin{align}
 U_{F}&=2\mu_0^2\Lambda_0\ell
    [w_F-\ell(1-e^{-w_F/\ell})],\nonumber\\
 U_{B}&=2\mu_0^2\Lambda_0\ell
    [w_B-\ell(1-e^{-w_B/\ell})],\nonumber\\
 U_{FB}&=\mu_0^2\Lambda_0\ell^2e^{-\eta_{\rm gap}/\ell}
    (1-e^{-w_F/\ell})(1-e^{-w_B/\ell}),
 \label{eq:exp_pair_integrals}
\end{align}
where $w_F,w_B$ are the window widths and $\eta_{\rm gap}\ge0$ is the nearest-edge gap. The distance between window centers is related to these parameters: $\Delta\eta=\eta_{\rm gap}+\left(w_F+w_B\right)/2$. The source cumulants are
\begin{align}
 R_{10}&=a=\mu_{0}w_{F},& R_{01}&=b=\mu_{0}w_{B},\nonumber\\
 R_{20}&=a+U_{F},&R_{02}&=b+U_{B},\nonumber\\
 R_{11}&=U_{FB},\nonumber\\
 R_{30}&=a+3U_{F},&R_{03}&=b+3U_{B},\nonumber\\
 R_{21}&=U_{FB},&R_{12}&=U_{FB},\nonumber\\
 R_{40}&=a+7U_{F},&R_{04}&=b+7U_{B},\nonumber\\
 R_{31}&=U_{FB},&R_{22}&=U_{FB},&R_{13}&=U_{FB},
 \label{eq:pair_ordinary_all}
\end{align}
and for low-order strongly intensive observables we get:
\begin{align}
 \Sigma&=1+\frac{b^2U_F+a^2U_B-2abU_{FB}}{ab(a+b)}.
 \label{eq:pair_aux_low}
\end{align}
\begin{align}
 \Delta&=1+\frac{b^2U_F-a^2U_B}{ab(b-a)}.
\end{align}
\begin{align}
 \Gamma&=1+\frac{3\bigl(b^3U_F-a^3U_B-ab(b-a)U_{FB}\bigr)}{ab(a+b)(b-a)}.
 \label{eq:pair_aux_high}
\end{align}
Appendix~\ref{app:vechernin} gives a compact evaluation prescription for the other $J_{ij}$ and their explicit symmetric limit. We choose the 900~GeV parameters from Ref.~\cite{AndronovVechernin2019}: $\mu_0\Lambda_0=0.73$ and $\ell=1.52$. Figure~\ref{fig:pair_av2019} shows results for three combinations of window widths: $(w_F,w_B)=(0.2,0.2)$, $(0.4,0.2)$ and $(0.4,0.05)$.

At fixed widths, $\Delta$ is independent of $\Delta\eta$ because it contains only within-window pair integrals. The normalizations of $\Delta$ and $\Gamma$ are undefined for equal means. In the exchange-symmetric case, several regular coordinates coincide with $\Sigma$, whereas $J_{21}$, $J_{22}$, and the fourth-order contrast contain additional nonlinear combinations of the pair integrals. Pair correlations alone therefore already separate the hierarchy from the conventional observables, without any genuine three- or four-particle dynamics.

\section{\label{sec:pythia}PYTHIA calculations}
\subsection{Three samples with a common generator version}
The analytical examples isolate particular contributions to the observables. PYTHIA implements a more detailed particle-production model in which these contributions occur together. We use three samples generated with PYTHIA 8.318 \cite{Pythia83}: Monash at $\sqrt{s}=900$ GeV, Trieste 2 with close packing at 900 GeV, and Monash at 13 TeV.

We compare the Monash results with the Trieste 2 tune, which incorporates string close-packing effects into PYTHIA~\cite{Altmann:2025afh,Bernardinis:2026hyp}. In this description, overlapping strings modify the effective color field experienced during fragmentation, making the string tension dependent on the surrounding string environment. This provides a phenomenological realization of the broader idea that overlapping strings need not fragment independently. A related physical motivation underlies the string-fusion model, in which overlapping color strings form sources with modified particle-production properties~\cite{BRAUN1993542}. The comparison is therefore of interest for testing how strongly intensive multiplicity observables respond to changes in string fragmentation associated with overlap. The two descriptions implement this idea differently, and the Monash–Trieste 2 comparison also includes other differences between the tunes. It should therefore not be interpreted as an isolated measurement of the close-packing contribution.

Each sample contains 150 million generated inelastic $ pp$ events. The two 900 GeV samples isolate the dependence on the generator setting at a common energy, while the Monash samples permit an energy comparison with the setting held fixed.

The generation uses \texttt{SoftQCD:inelastic=on} with proton beams and \texttt{Tune:pp=14}. Trieste 2 is applied on top of this baseline, including its colour-reconnection and close-packing settings \cite{Altmann:2025afh}. Appendix~\ref{app:production} lists the complete override card and the corresponding Monash values.

INEL$>0$ requires at least one charged final particle in $|\eta|<1$, with no imposed transverse-momentum threshold. The quoted 150 million precede the INEL$>0$ selection. The V0 activity estimator is $V=N_{\rm A}+N_{\rm C}$, where $N_{\rm A}$ counts charged final particles in $2.8<\eta<5.1$ and $N_{\rm C}$ in $-3.7<\eta<-1.7$, again without a $p_{\rm T}$ threshold and over full azimuth. It is a multiplicity proxy for the V0 acceptance of the ALICE experiment~\cite{ALICE:2013axi}, not a simulated detector amplitude.

The measurement windows, in contrast, count particles with $0.2<p_{\mathrm T}<2$ GeV. Unless specified otherwise they cover full azimuth and are defined in pseudorapidity. Their center separation is $\Delta\eta=\eta_F-\eta_B$, their full widths are $w_F,w_B$, and their nearest-edge separation is $\eta_{\rm gap}=\Delta\eta-(w_F+w_B)/2$ for disjoint windows.

Statistical uncertainties are estimated using the jackknife method~\cite{Efron1981TheJE}, dividing each sample into 150 equal-sized blocks and omitting one block at a time. These errors describe event sampling, not generator or detector systematics.

\subsection{Scans over window geometry and collision energy}
\begin{figure*}[!t]
\centering
\includegraphics[width=\textwidth]{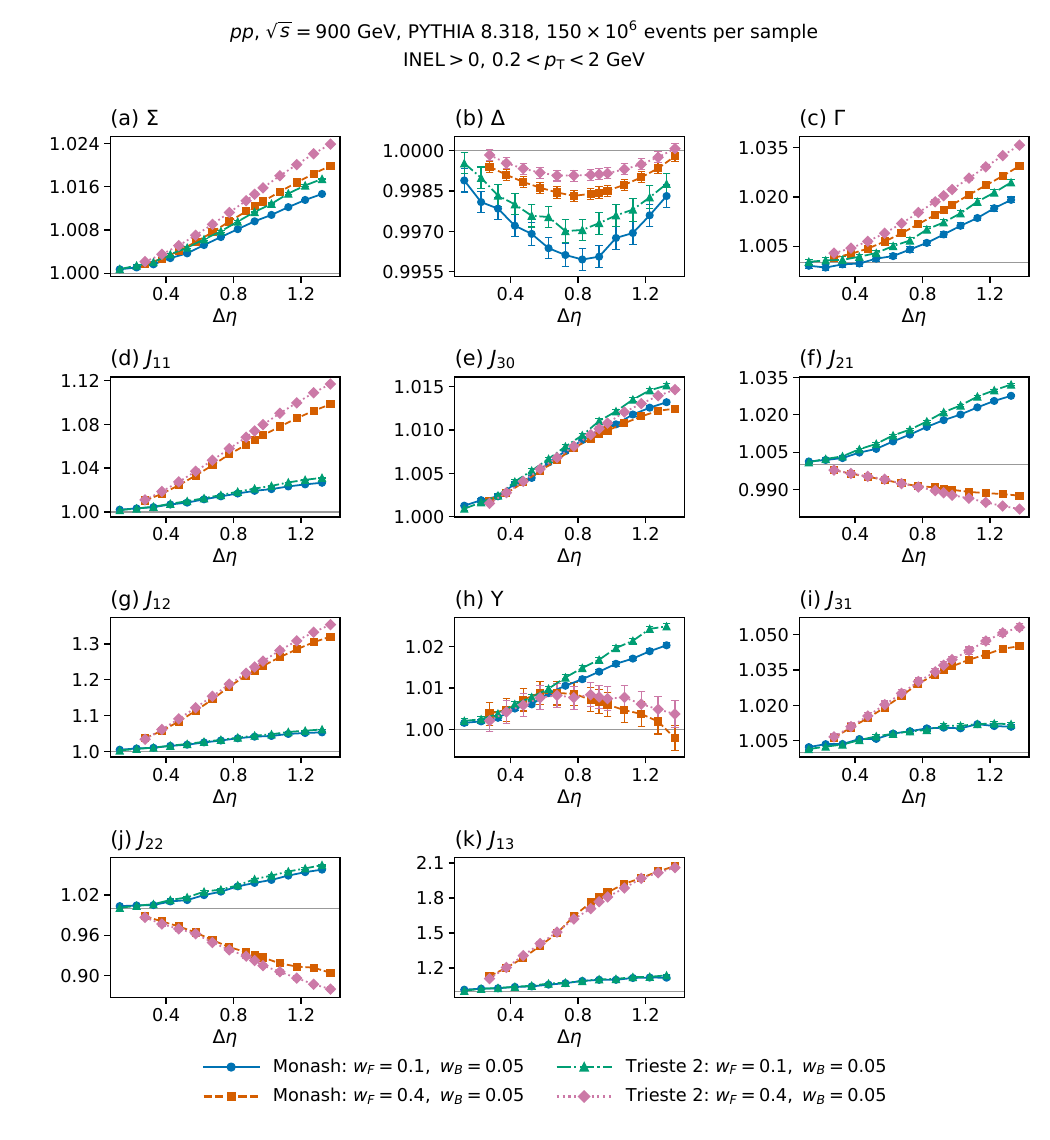}
\caption{Pseudorapidity scan: Monash and Trieste 2 PYTHIA 8.318 at $ pp$, $\sqrt{s}=900$ GeV, 150 million events per sample. INEL$>0$, $0.2<p_{\mathrm T}<2$ GeV. The backward window has $\eta_B=-0.775$ and width 0.05. Forward widths are 0.1 and 0.4. Panels (a)--(k) show $\Sigma$, $\Delta$, $\Gamma$, $J_{11}$, $J_{30}$, $J_{21}$, $J_{12}$, $\Upsilon$, $J_{31}$, $J_{22}$, and $J_{13}$, respectively. Bars show jackknife standard errors.}
\label{fig:pythiascan8318}
\end{figure*}

\begin{figure*}[!t]
\centering
\includegraphics[width=\textwidth]{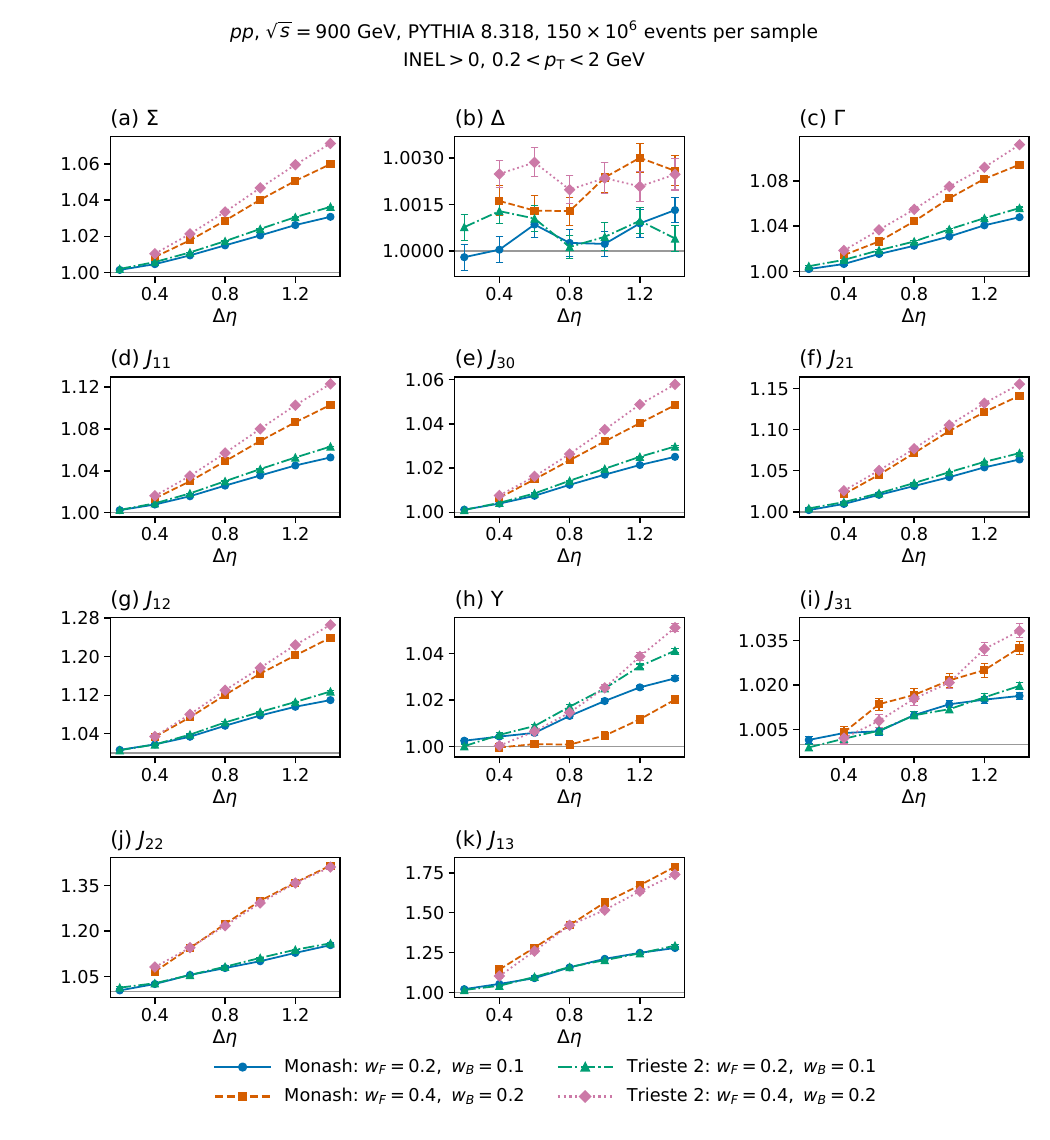}
\caption{Monash/Trieste 2 comparison in PYTHIA 8.318 at $ pp$, $\sqrt{s}=900$ GeV, 150 million events per sample. INEL$>0$, $0.2<p_{\mathrm T}<2$ GeV, pseudorapidity-width pairs $0.2/0.1$ and $0.4/0.2$, centers at $\pm \Delta\eta/2$, and $\Delta\eta\leq1.4$. Panels (a)--(k) show $\Sigma$, $\Delta$, $\Gamma$, $J_{11}$, $J_{30}$, $J_{21}$, $J_{12}$, $\Upsilon$, $J_{31}$, $J_{22}$, and $J_{13}$, respectively. Bars show jackknife standard errors.}
\label{fig:windows8318}
\end{figure*}

The scans probe how the observables depend on the position and widths of the two windows. In the geometry of Fig.~\ref{fig:pythiascan8318}, the backward window is fixed at $\eta_B=-0.775$ with $w_B=0.05$, and the forward window with $w_F=0.1$ or $0.4$ moves through the allowed separations, $\Delta\eta\le1.4$. In the geometry of Fig.~\ref{fig:windows8318}, the windows are centered at $\pm\Delta\eta/2$ with width pairs $0.2/0.1$ and $0.4/0.2$. Both scans at 900~GeV show the same qualitative pattern: $\Sigma$ increases with separation, while the higher-order coordinates need not follow it. The response also changes its magnitude and shape when the widths vary, because the window integrals weight the underlying correlations differently.

The Monash and Trieste~2 curves in these figures use the same acceptance and event selection. The differences between the curves therefore reflect the specified settings within PYTHIA~8.318 rather than the generator release. The Trieste~2 tune gives larger $\Sigma$ than Monash, a shift in the same direction as that found for the string-fusion mechanism modeled in Ref.~\cite{Belokurova:2022lcr}. The higher-order coordinates also respond to the tune change, which does not isolate the contribution of close packing.

\begin{figure*}[!t]
\centering
\includegraphics[width=\textwidth]{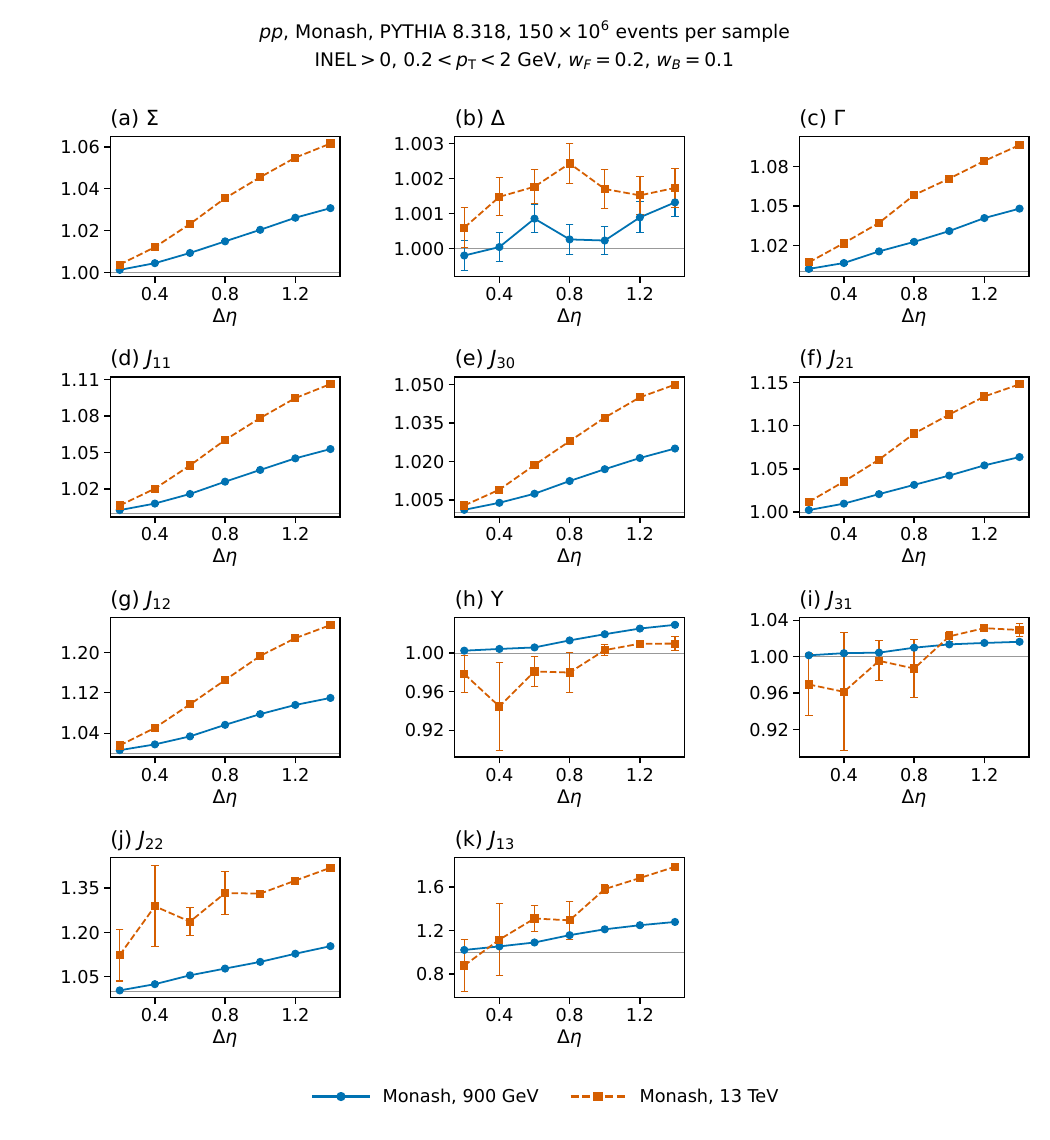}
\caption{Energy comparison: Monash 8.318 at $\sqrt{s}=900$ GeV and 13 TeV, $ pp$, 150 million events per sample. Inclusive INEL$>0$, pseudorapidity widths $0.2/0.1$, $0.2<p_{\mathrm T}<2$ GeV. Panels (a)--(k) show $\Sigma$, $\Delta$, $\Gamma$, $J_{11}$, $J_{30}$, $J_{21}$, $J_{12}$, $\Upsilon$, $J_{31}$, $J_{22}$, and $J_{13}$, respectively. Bars show jackknife standard errors.}
\label{fig:energy8318}
\end{figure*}

Figure~\ref{fig:energy8318} compares the two collision energies with the same asymmetric pseudorapidity windows as in Fig.~\ref{fig:windows8318}. Changing $\sqrt{s}$ modifies the magnitude of $J_{ij}$ but not its dependence on $\Delta\eta$.

\subsection{Classes selected with the V0M-like activity estimator}
\begin{figure*}[!t]
\centering
\includegraphics[width=\textwidth]{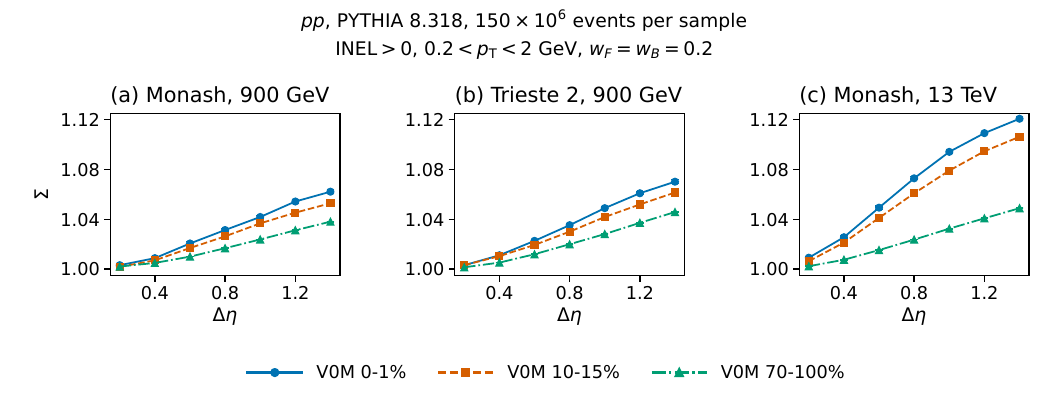}
\caption{Class dependence of $\Sigma$ for Monash and Trieste 2 at 900 GeV and Monash at 13 TeV, PYTHIA 8.318, 150 million events per sample. Activity percentiles are defined within each sample's INEL$>0$ population using the V0M-like estimator of Sec.~\ref{sec:pythia}. Symmetric pseudorapidity windows have width 0.2, and $0.2<p_{\mathrm T}<2$ GeV. Panels (a), (b), and (c) correspond to Monash at 900 GeV, Trieste 2 at 900 GeV, and Monash at 13 TeV, respectively. Bars show jackknife standard errors.}
\label{fig:sigma_classes8318}
\end{figure*}

\begin{figure*}[!t]
\centering
\includegraphics[width=\textwidth]{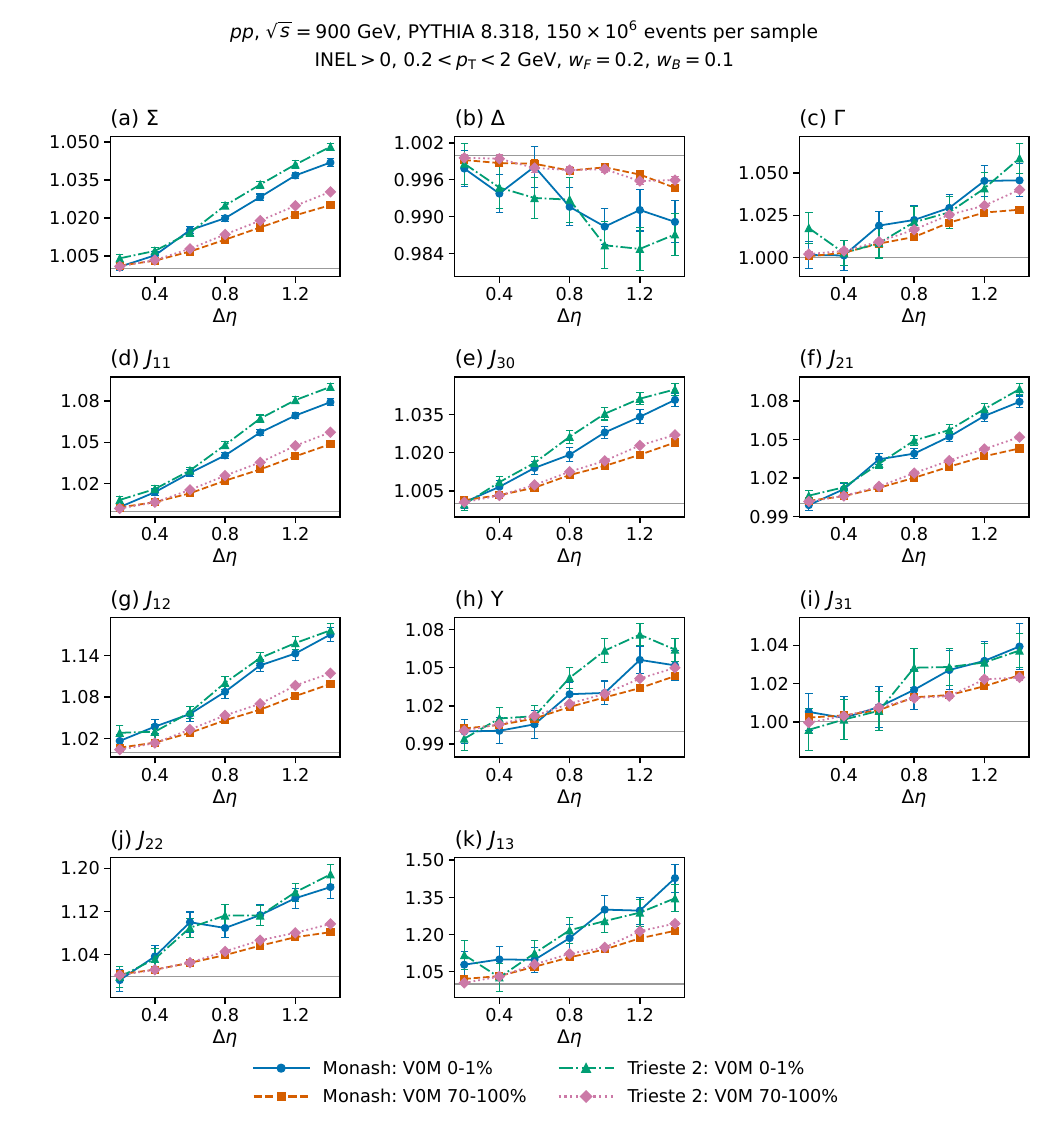}
\caption{Class-dependent comparison of Monash and Trieste 2, PYTHIA 8.318, in $pp$ collisions at $\sqrt{s}=900$ GeV, 150 million events per sample. Activity percentiles refer to each sample's INEL$>0$ population. Pseudorapidity widths are $0.2$ and $0.1$, centers are $\pm \Delta\eta/2$, and $0.2<p_{\mathrm T}<2$ GeV. Panels (a)--(k) show $\Sigma$, $\Delta$, $\Gamma$, $J_{11}$, $J_{30}$, $J_{21}$, $J_{12}$, $\Upsilon$, $J_{31}$, $J_{22}$, and $J_{13}$, respectively. Bars show jackknife standard errors.}
\label{fig:classbasis8318}
\end{figure*}

We evaluate the observables in percentile classes of the activity estimator $V$, defined separately for each sample within its INEL$>0$ population. Figure~\ref{fig:sigma_classes8318} shows $\Sigma$ in symmetric windows, while Fig.~\ref{fig:classbasis8318} compares the regular higher-order basis in asymmetric windows at 900 GeV.

$\Sigma$ increases with the window separation in every class and sample, and the classes are ordered by activity: the 0--1\% class lies highest and the 70--100\% class lowest in every panel. The class separation grows with the separation. At 900~GeV the curves of all classes meet at $\Sigma\approx1$ for small separations and reach $\Sigma-1\approx0.06$ in the 0--1\% class at $\Delta\eta=1.4$, with a class spread of about $0.02$--$0.03$. At 13~TeV the rise is twice as large, $\Sigma-1\approx0.12$ in the 0--1\% class, and the class separation is already visible at the smallest separation. The tune effect is smaller: Trieste~2 lies slightly above Monash at a fixed class and separation.

Across much of the higher-order basis at 900~GeV (Fig.~\ref{fig:classbasis8318}), the 0--1\% class lies above the 70--100\% class, while the tune changes generally produce smaller shifts. This ordering is not universal for every coordinate and separation. Several higher-order points overlap within their statistical errors or reverse the tune ordering. The observable $\Delta$ stays close to unity, but some class-dependent departures are resolved by the displayed statistical errors.

Selection on V0 activity can alter both the source-count distribution and the emission law itself. Strongly intensive observables cancel only the former, and only while all sources share a common emission law. A remaining class dependence therefore reflects a change in the source composition, in the source emission, or in correlations induced by the selection. The class scans alone cannot separate these mechanisms.

\subsection{Joint pseudorapidity and azimuthal separations}
Azimuthal separation resolves structure that the full-azimuth scans average over. To expose it, we split the azimuth into two sectors of width $\pi/4$, keep the pseudorapidity widths $w_F=0.4$ and $w_B=0.2$, and evaluate the observables as surfaces over the window-center separations $\Delta\eta$ and $\Delta\phi$.

\begin{figure*}[!t]
\centering
\includegraphics[width=\textwidth]{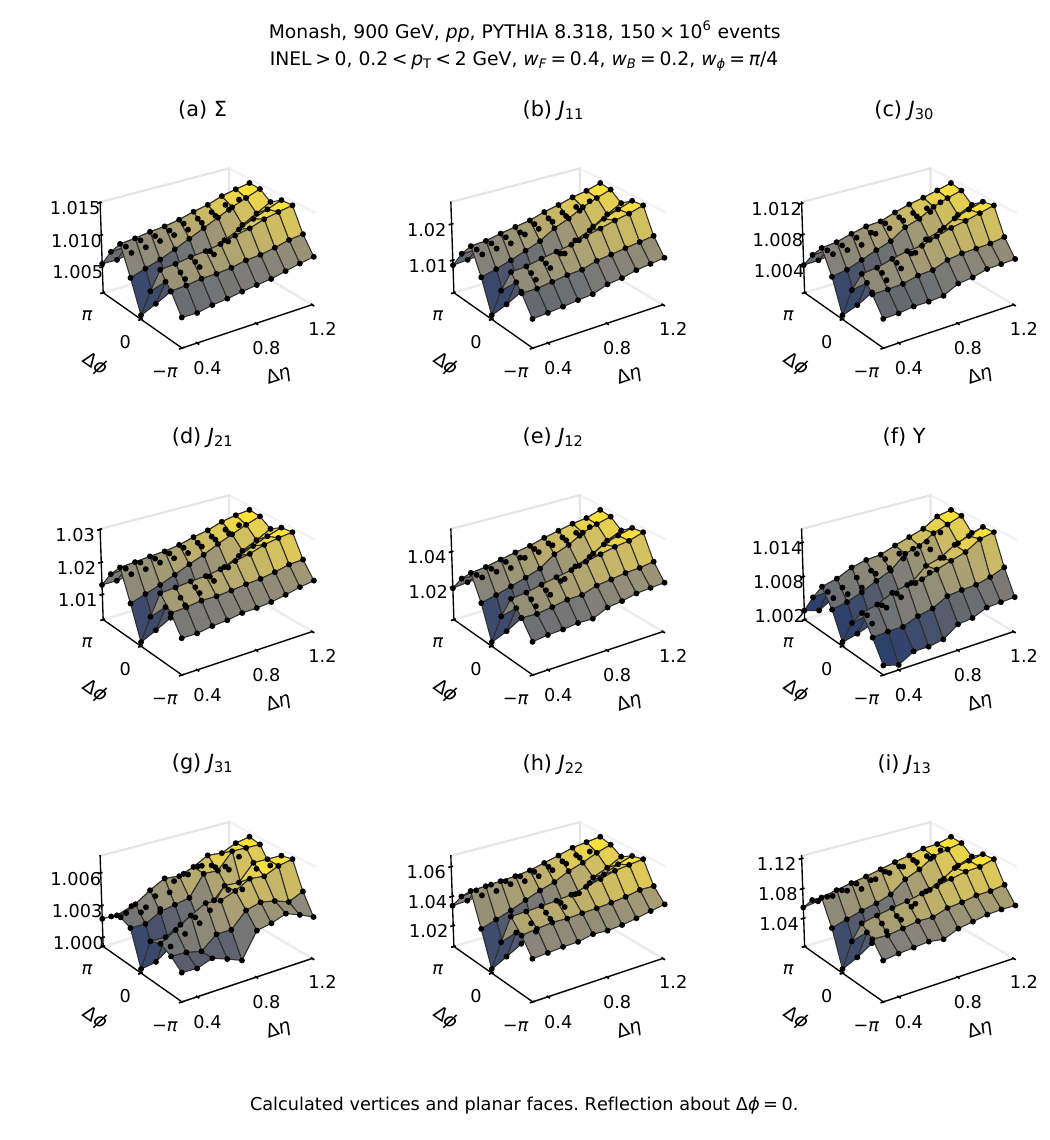}
\caption{PYTHIA 8.318 Monash results for $ pp$, $\sqrt{s}=900$ GeV, 150 million events per sample, inclusive INEL$>0$, $0.2<p_{\mathrm T}<2$ GeV. Pseudorapidity widths are $0.4/0.2$ and both azimuthal widths are $\pi/4$. The surfaces are shown over the window-center separations $\Delta\eta$ and $\Delta\phi$. Surface vertices are the evaluated geometries. Flat connecting faces guide the eye without smoothing. The calculated half-plane $0\le\Delta\phi\le\pi$ is reflected about $\Delta\phi=0$ for presentation. Panels (a)--(i) show $\Sigma$, $J_{11}$, $J_{30}$, $J_{21}$, $J_{12}$, $\Upsilon$, $J_{31}$, $J_{22}$, and $J_{13}$, respectively. Color follows surface height with a separate scale in each panel.}
\label{fig:phi8318_monash900}
\end{figure*}

\begin{figure*}[!t]
\centering
\includegraphics[width=\textwidth]{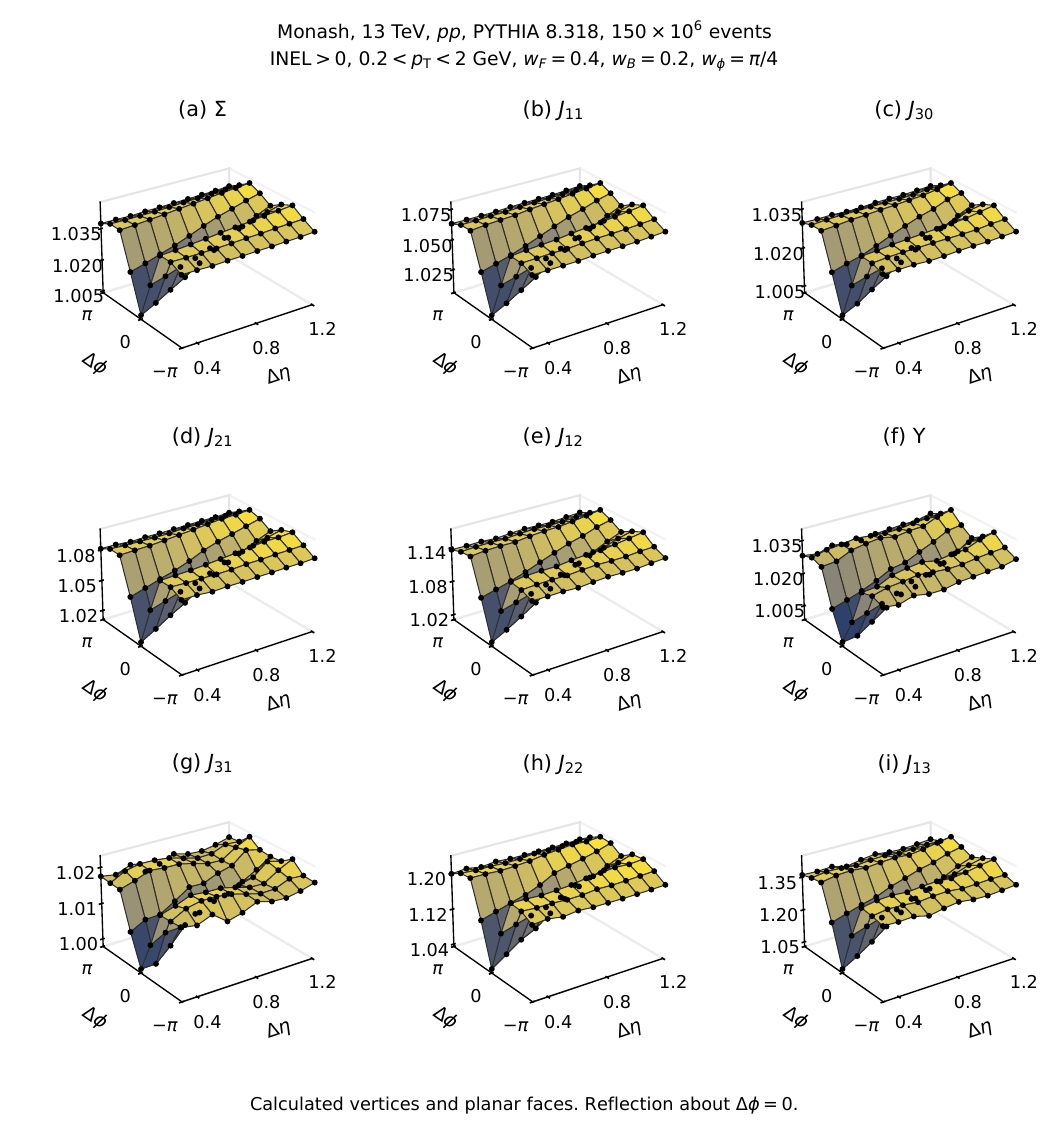}
\caption{PYTHIA 8.318 Monash results for $ pp$, $\sqrt{s}=13$ TeV, 150 million events per sample, inclusive INEL$>0$, $0.2<p_{\mathrm T}<2$ GeV. Pseudorapidity widths are $0.4/0.2$ and both azimuthal widths are $\pi/4$. The surfaces are shown over the window-center separations $\Delta\eta$ and $\Delta\phi$. Surface vertices are the evaluated geometries. Flat connecting faces guide the eye without smoothing. The calculated half-plane $0\le\Delta\phi\le\pi$ is reflected about $\Delta\phi=0$ for presentation. Panels (a)--(i) show $\Sigma$, $J_{11}$, $J_{30}$, $J_{21}$, $J_{12}$, $\Upsilon$, $J_{31}$, $J_{22}$, and $J_{13}$, respectively. Color follows surface height with a separate scale in each panel.}
\label{fig:phi8318_monash13000}
\end{figure*}

Figure~\ref{fig:phi8318_monash900} shows the Monash maps at 900~GeV and Fig.~\ref{fig:phi8318_monash13000} at 13~TeV for the displayed strongly intensive coordinates as functions of $\Delta\eta$ and $\Delta\phi$. $\Delta$ and $\Gamma$ are omitted for brevity. The Trieste~2 maps at 900~GeV are omitted as well: they are visually indistinguishable from the Monash ones, and the tune effect is quantified by the full-azimuth scans of Figs.~\ref{fig:pythiascan8318} and~\ref{fig:windows8318}.

All surfaces rise with $\Delta\eta$, and the rise is steepest at $\Delta\phi=0$, flattening toward $\Delta\phi=\pi$: the dependence on the two separations does not factorize. At small $\Delta\eta$ the surfaces dip toward unity and locally below it for intermediate $\Delta\phi$. The amplitudes follow the hierarchy already seen in the one-dimensional scans. At 13~TeV the surfaces flatten in $\Delta\phi$ once $\Delta\eta$ is large, so the azimuthal modulation is an effect of the correlated region, not a uniform offset.

The surfaces connect the evaluated geometries without smoothing, and the calculated half-plane $0\le\Delta\phi\le\pi$ is reflected about $\Delta\phi=0$ for display. Statistical errors are not drawn on the surfaces. The full-azimuth errors in Figs.~\ref{fig:pythiascan8318}--\ref{fig:energy8318} do not quantify the uncertainties for these narrower azimuthal sectors, so the surfaces are used here to illustrate qualitative trends.

\subsection{\label{sec:pythia_fit}Fitting the density and testing a pair-only closure}
We test whether a specified single-source density and pair correlation function can jointly describe the PYTHIA mean multiplicities, $\Sigma$, and $\Delta$, and predict the higher-order observables. As in Sec.~\ref{sec:definitions}, all sources are assumed to share one emission law. The measured one-particle density therefore constrains the source-profile shape, but determines only the product of the mean source count and the single-source normalization.

We use the even profile
\begin{equation}
 f(\eta)=1+c_2\eta^2+c_4\eta^4,
 \qquad
 \lambda_{1}(\eta)=\mu_{0} f(\eta),
\end{equation}
as a local parametrization in the central region $|\eta|\leq0.8$ containing the fitted windows. The corresponding event density is $\rho_{\mathrm{event}}(\eta)=\rho_0f(\eta)$, with $\rho_0=\langle S\rangle\mu_{0}$.

For the connected pair correlation we adopt the Gaussian form
\begin{equation}
 \Lambda(\eta_1,\eta_2)=\Lambda_0
 \exp\!\left[-\frac{(\eta_1-\eta_2)^2}{2\ell^2}\right].
 \label{eq:gaussian_kernel}
\end{equation}
This is a separate model assumption from the exponential kernel considered in Sec.~\ref{sec:models}. The comparison below tests this specified Gaussian model.

The joint fit uses windows of widths $0.2$ and $0.1$, centered at $\eta_F=\Delta\eta/2$ and $\eta_B=-\Delta\eta/2$. Its inputs are $\langle n_F\rangle$, $\langle n_B\rangle$, $\Sigma$, and $\Delta$, evaluated at different window-center separations $\Delta\eta$. The five fitted parameters are $\rho_0$, $c_2$, $c_4$, $a_2=\mu_{0}\Lambda_0$, and $\ell$.

\begin{figure*}[!t]
\centering
\includegraphics[width=\textwidth]{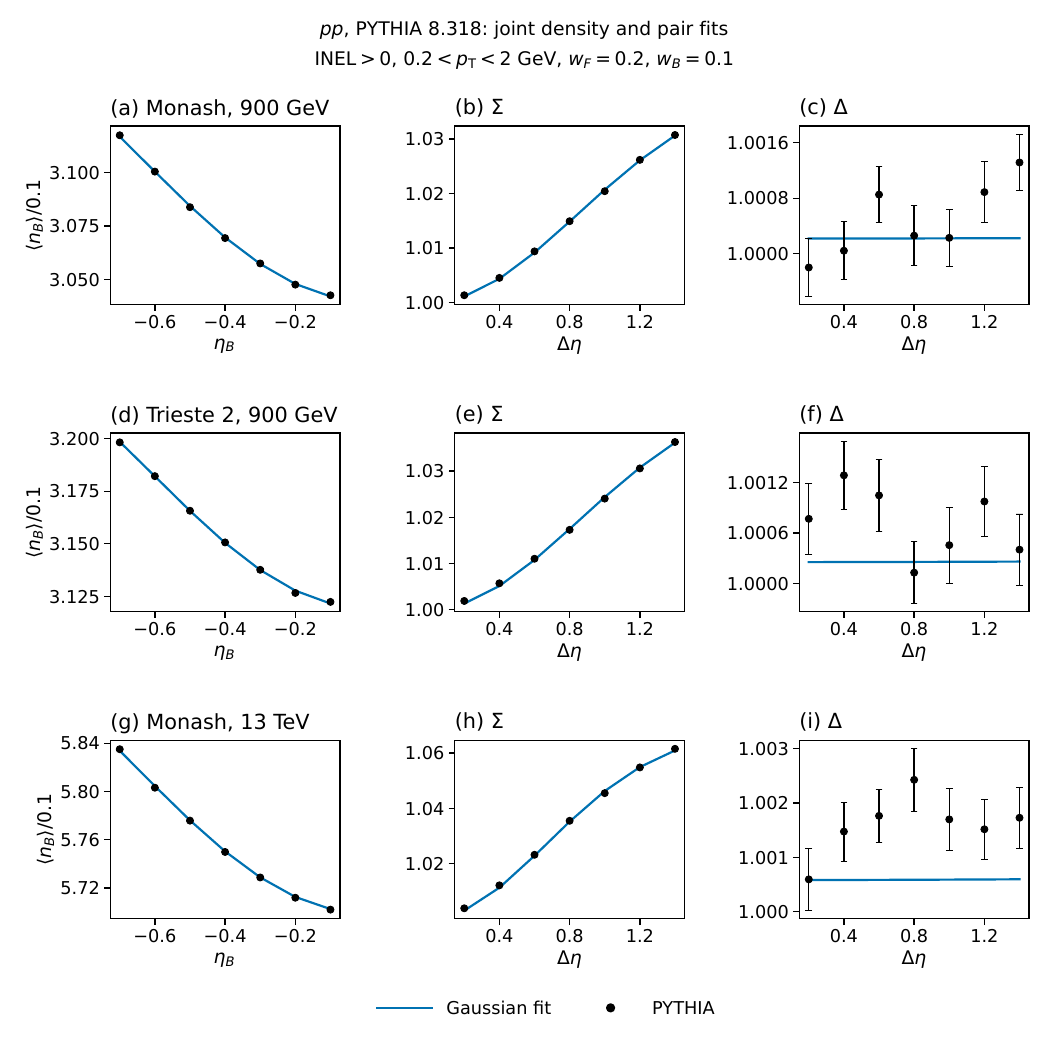}
\caption{Joint Gaussian fit to PYTHIA 8.318 mean multiplicities, $\Sigma$, and $\Delta$ for $pp$ collisions. Black points denote PYTHIA with jackknife standard errors and blue curves the model. Each sample contains 150 million generated events with INEL$>0$ selection and $0.2<p_{\mathrm T}<2$ GeV. Window widths are $0.2/0.1$. Panels (a)--(c), (d)--(f), and (g)--(i) correspond to Monash at 900 GeV, Trieste 2 at 900 GeV, and Monash at 13 TeV. Columns show $\langle n_B\rangle/0.1$ at $\eta_B=-\Delta\eta/2$, $\Sigma$, and $\Delta$. Only the $\Delta$ points with $\Delta\eta\leq1.0$ enter the fit.}
\label{fig:profile_joint8318}
\end{figure*}

Figure~\ref{fig:profile_joint8318} displays the fits, and Table~\ref{tab:lowerfits} lists their parameters. The reported PYTHIA $\chi^2$ values use the full paired jackknife covariance matrix of the inputs included in each fit, with $\Delta$ restricted to $\Delta\eta\leq1.0$ in the lower-order fit, and are treated as nominal fit diagnostics. The density panels show the finite-window average $\langle n_B\rangle/0.1$ at the backward-window center.

\begin{table*}[t]
\caption{Parameters of the joint density and pair fit. The identifiable normalization combinations are $\rho_0=\langle S\rangle\mu_{0}$ and $a_2=\mu_{0}\Lambda_0$.}
\label{tab:lowerfits}
\begin{ruledtabular}
\begin{tabular}{lccc}
 & Monash 900 GeV & Trieste 2 900 GeV & Monash 13 TeV\\
$\rho_0$
 & $3.0400\pm0.0003$
 & $3.1194\pm0.0003$
 & $5.6989\pm0.0005$\\
$c_2$
 & $0.0653\pm0.0008$
 & $0.0672\pm0.0007$
 & $0.0602\pm0.0006$\\
$c_4$
 & $-0.028\pm0.001$
 & $-0.031\pm0.001$
 & $-0.024\pm0.001$\\
$a_2$
 & $0.294\pm0.004$
 & $0.349\pm0.004$
 & $0.499\pm0.002$\\
$\ell$
 & $0.82\pm0.01$
 & $0.825\pm0.008$
 & $0.651\pm0.003$\\
$\chi^2/\mathrm{ndf}$
 & $23.82/21$
 & $45.93/21$
 & $163.24/21$\\
\end{tabular}
\end{ruledtabular}
\end{table*}

\begin{figure*}[!t]
\centering
\includegraphics[width=\textwidth]{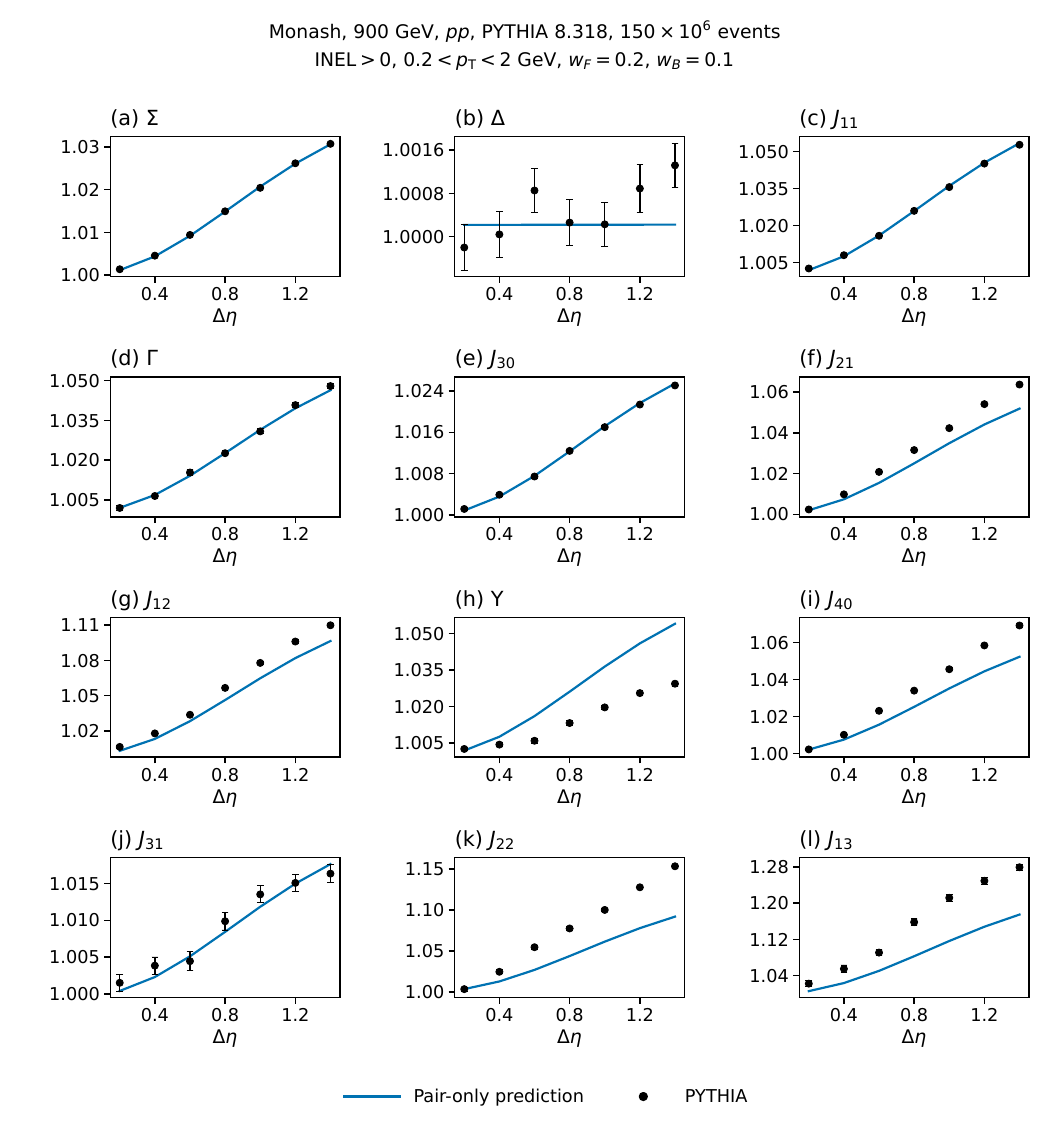}
\caption{Pair-only predictions (curves) compared with PYTHIA 8.318 results (points) for Monash at $\sqrt{s}=900~\mathrm{GeV}$, using 150 million generated $pp$ events. The selection is INEL$>0$ with $0.2<p_{\mathrm T}<2$ GeV and window widths $0.2/0.1$. Bars show jackknife standard errors. Lower-order parameters are held fixed and their uncertainties are not propagated into bands. Panels (a)--(l) show $\Sigma$, $\Delta$, $J_{11}$, $\Gamma$, $J_{30}$, $J_{21}$, $J_{12}$, $\Upsilon$, $J_{40}$, $J_{31}$, $J_{22}$, and $J_{13}$, respectively.}
\label{fig:profile_higher8318_monash900}
\end{figure*}

\begin{figure*}[!t]
\centering
\includegraphics[width=\textwidth]{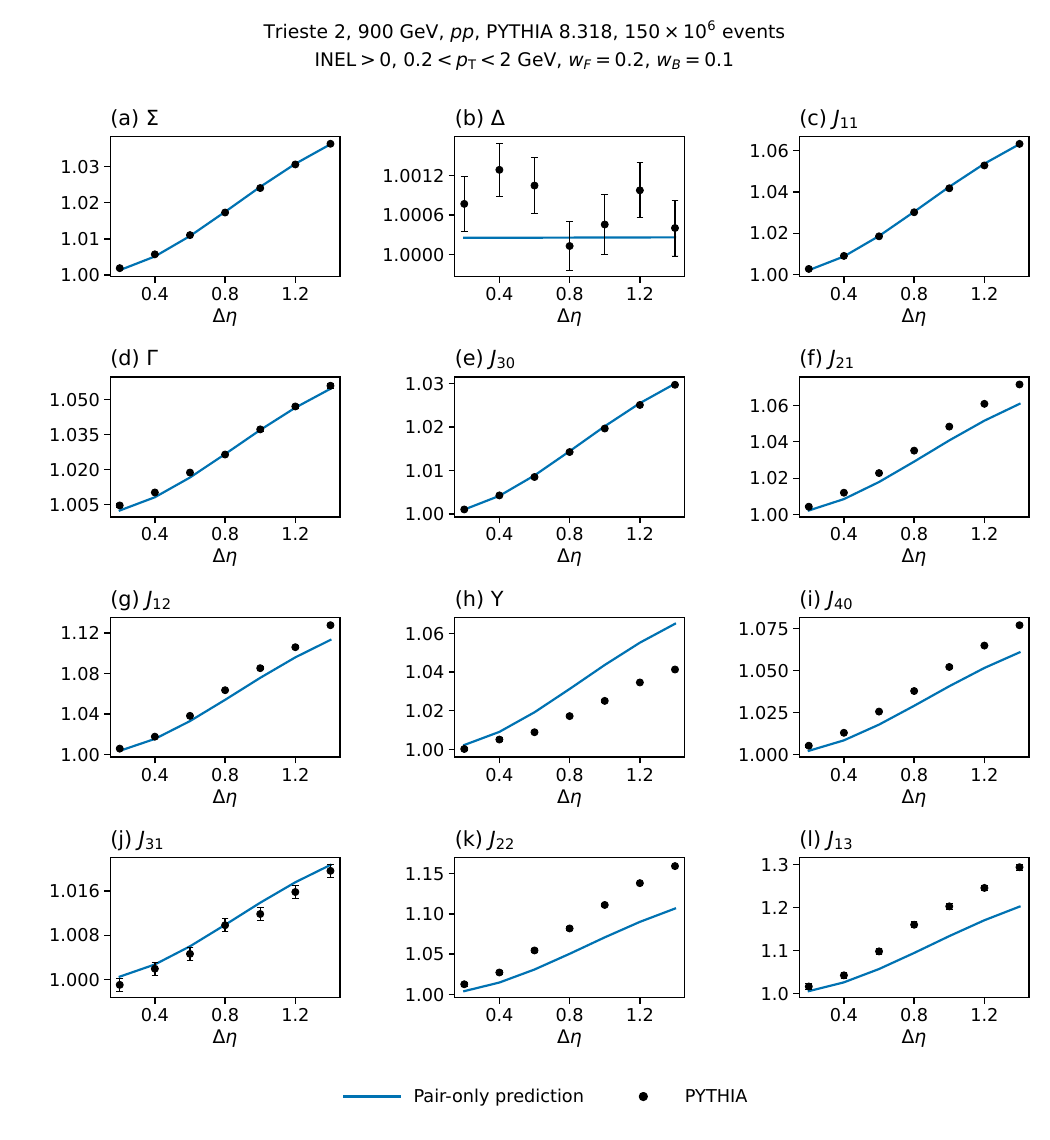}
\caption{Pair-only predictions (curves) compared with PYTHIA 8.318 results (points) for Trieste~2 at $\sqrt{s}=900~\mathrm{GeV}$, using 150 million generated $pp$ events. The selection is INEL$>0$ with $0.2<p_{\mathrm T}<2$ GeV and window widths $0.2/0.1$. Bars show jackknife standard errors. Lower-order parameters are held fixed and their uncertainties are not propagated into bands. Panels (a)--(l) show $\Sigma$, $\Delta$, $J_{11}$, $\Gamma$, $J_{30}$, $J_{21}$, $J_{12}$, $\Upsilon$, $J_{40}$, $J_{31}$, $J_{22}$, and $J_{13}$, respectively.}
\label{fig:profile_higher8318_trieste900}
\end{figure*}

\begin{figure*}[!t]
\centering
\includegraphics[width=\textwidth]{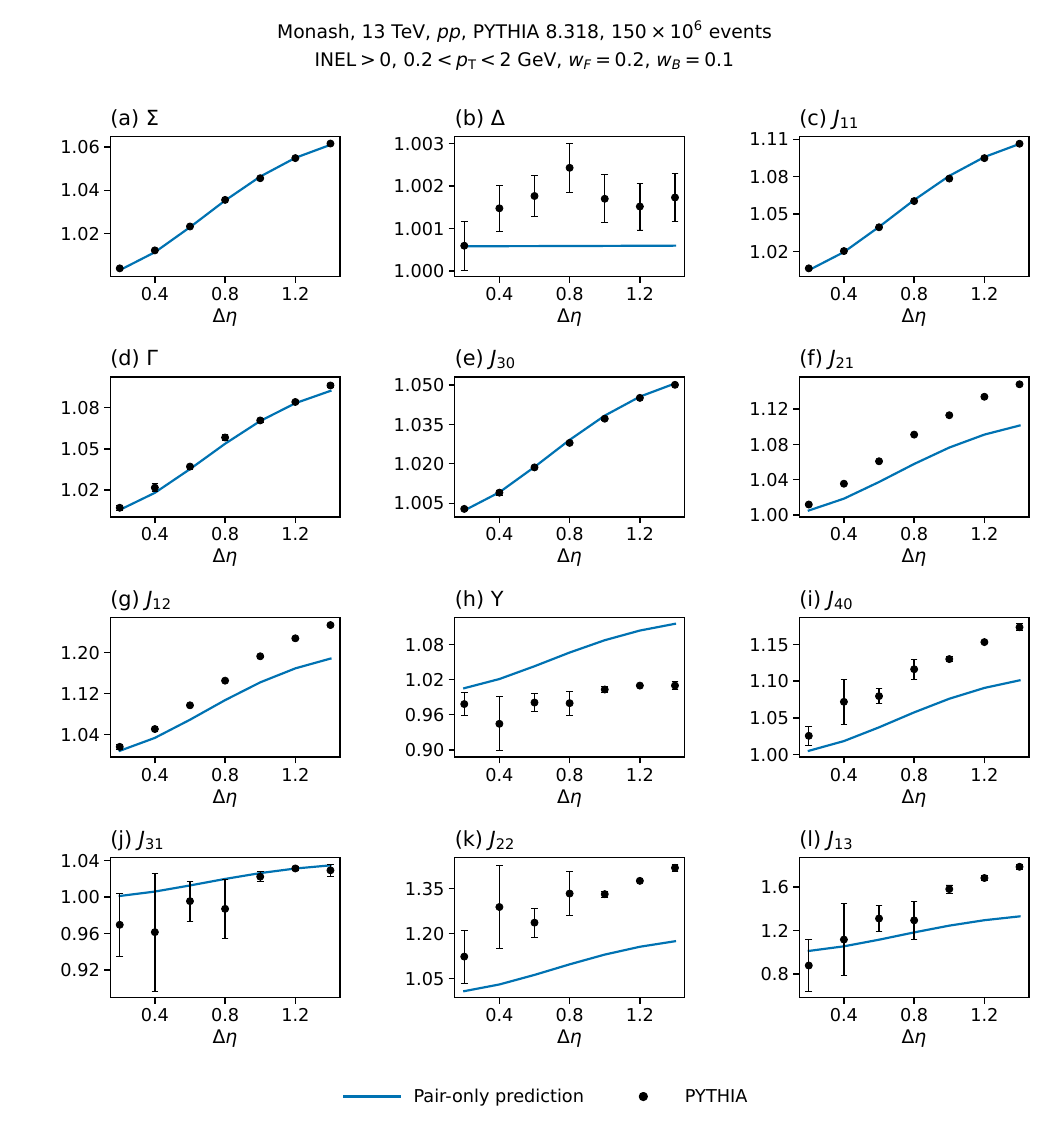}
\caption{Pair-only predictions (curves) compared with PYTHIA 8.318 results (points) for Monash at $\sqrt{s}=13~\mathrm{TeV}$, using 150 million generated $pp$ events. The selection is INEL$>0$ with $0.2<p_{\mathrm T}<2$ GeV and window widths $0.2/0.1$. Bars show jackknife standard errors. Lower-order parameters are held fixed and their uncertainties are not propagated into bands. Panels (a)--(l) show $\Sigma$, $\Delta$, $J_{11}$, $\Gamma$, $J_{30}$, $J_{21}$, $J_{12}$, $\Upsilon$, $J_{40}$, $J_{31}$, $J_{22}$, and $J_{13}$, respectively.}
\label{fig:profile_higher8318_monash13000}
\end{figure*}

With the fitted density and pair kernel held fixed, setting the connected three- and four-particle densities to zero determines all observables through fourth order without additional parameters. Figures~\ref{fig:profile_higher8318_monash900}--\ref{fig:profile_higher8318_monash13000} compare these pair-only predictions with PYTHIA. The three comparisons use the same $0.2/0.1$ windows and center separations up to $1.4$, with INEL$>0$ event selection and $0.2<p_{\mathrm T}<2$~GeV. Points carry jackknife standard errors. The curves are conditional central predictions: uncertainty in the fitted lower-order parameters has not been propagated into prediction bands.

These comparisons test the higher-order consequences of the specified density and pair kernel. A discrepancy may reflect nonzero connected three- or four-particle correlations, an inadequate lower-order parametrization, or both. In particular, the substantial lower-order residuals for Monash at 13~TeV prevent attributing its higher-order discrepancies exclusively to missing connected correlations. The next subsection tests the effect of adding specified three- and four-particle terms while retaining this lower-order baseline.

\subsection{\label{sec:pythia_connected}Conditional fits of connected three- and four-particle terms}
We extend the pair-only model by introducing connected three- and four-particle densities with two additional amplitudes. The lower-order parameters are held fixed at the values in Table~\ref{tab:lowerfits}. We adopt the symmetric Gaussian hierarchy
\begin{align}
 C_n(\eta_1,\ldots,\eta_n)
 &=\mu_0 a_n\prod_{r=1}^{n}f(\eta_r)\nonumber\\
 &\quad\times\exp\!\left[-\frac{\sum_{r<s}(\eta_r-\eta_s)^2}{n\ell^2}\right],
 \label{eq:connected_gaussian}
\end{align}
where $f(\eta)=1+c_2\eta^2+c_4\eta^4$ is the previously fitted one-particle profile.

For $n=2$, Eq.~\eqref{eq:connected_gaussian} is consistent with the normalized pair kernel through
\begin{equation}
 C_2(\eta_1,\eta_2)
 =\lambda_1(\eta_1)\lambda_1(\eta_2)
 \Lambda(\eta_1,\eta_2),
 \qquad
 \lambda_1(\eta)=\mu_0 f(\eta),
\end{equation}
with $a_2=\mu_0\Lambda_0$. The same fitted length $\ell$ is used at all orders. No independent three- or four-particle correlation length is introduced. The higher-order calculation uses $\mu_0=1$ as a normalization convention, since the common factor $\mu_0$ in the connected densities cancels from the strongly intensive observables of this hierarchy. The explicit finite-window integrations, conversion to ordinary source cumulants, and evaluation of the strongly intensive coordinates for this Gaussian hierarchy are provided in Appendix~\ref{app:gaussian_explicit}.

The amplitudes are fitted sequentially. First, $a_3$ is determined from $J_{30}$, $J_{21}$, and $J_{12}$. The fitted $a_3$ is then held fixed while $a_4$ is determined from $\Upsilon$, $J_{31}$, $J_{22}$, and $J_{13}$.

Table~\ref{tab:higherfits} reports the fitted amplitudes and nominal fit diagnostics. The quoted amplitude errors condition on all preceding fit parameters. They do not include uncertainty propagated from earlier stages and are not errors from a simultaneous fit of the entire hierarchy.

\begin{table*}[t]
\caption{Conditional higher-order amplitudes and nominal fit diagnostics. Each arrow compares the objective with the new amplitude fixed to zero and with that amplitude fitted. The third-order baseline is the pair-only model. The fourth-order baseline already includes the fitted $a_3$, which remains fixed when fitting $a_4$.}
\label{tab:higherfits}
\begin{ruledtabular}
\begin{tabular}{lccc}
 & Monash 900 GeV & Trieste 2 900 GeV & Monash 13 TeV\\
$a_3$
 & $0.64\pm0.01$
 & $0.61\pm0.01$
 & $2.35\pm0.02$\\
$\chi^2_3$ ($\mathrm{ndf}=20$ after fit)
 & $2285.7\to35.5$
 & $2428.0\to51.3$
 & $11463.76\to65.06$\\
$a_4$
 & $3.3\pm0.3$
 & $2.3\pm0.3$
 & $10.7\pm0.8$\\
$\chi^2_4$ ($\mathrm{ndf}=27$ after fit)
 & $167.7\to51.9$
 & $105.1\to49.7$
 & $230.1\to52.8$\\
\end{tabular}
\end{ruledtabular}
\end{table*}

Adding a single three-particle amplitude reduces the nominal third-order objective by more than an order of magnitude in every sample. The fitted amplitudes are similar for the two tunes at 900~GeV. For Monash, $a_3$ increases by a factor of approximately 3.7 between 900~GeV and 13~TeV. This comparison refers to the specified hierarchy, with the profile and length fixed separately for each sample. With $a_3$ retained, fitting $a_4$ further reduces the fourth-order objective in all three samples.

\begin{figure*}[!t]
\centering
\includegraphics[width=\textwidth]{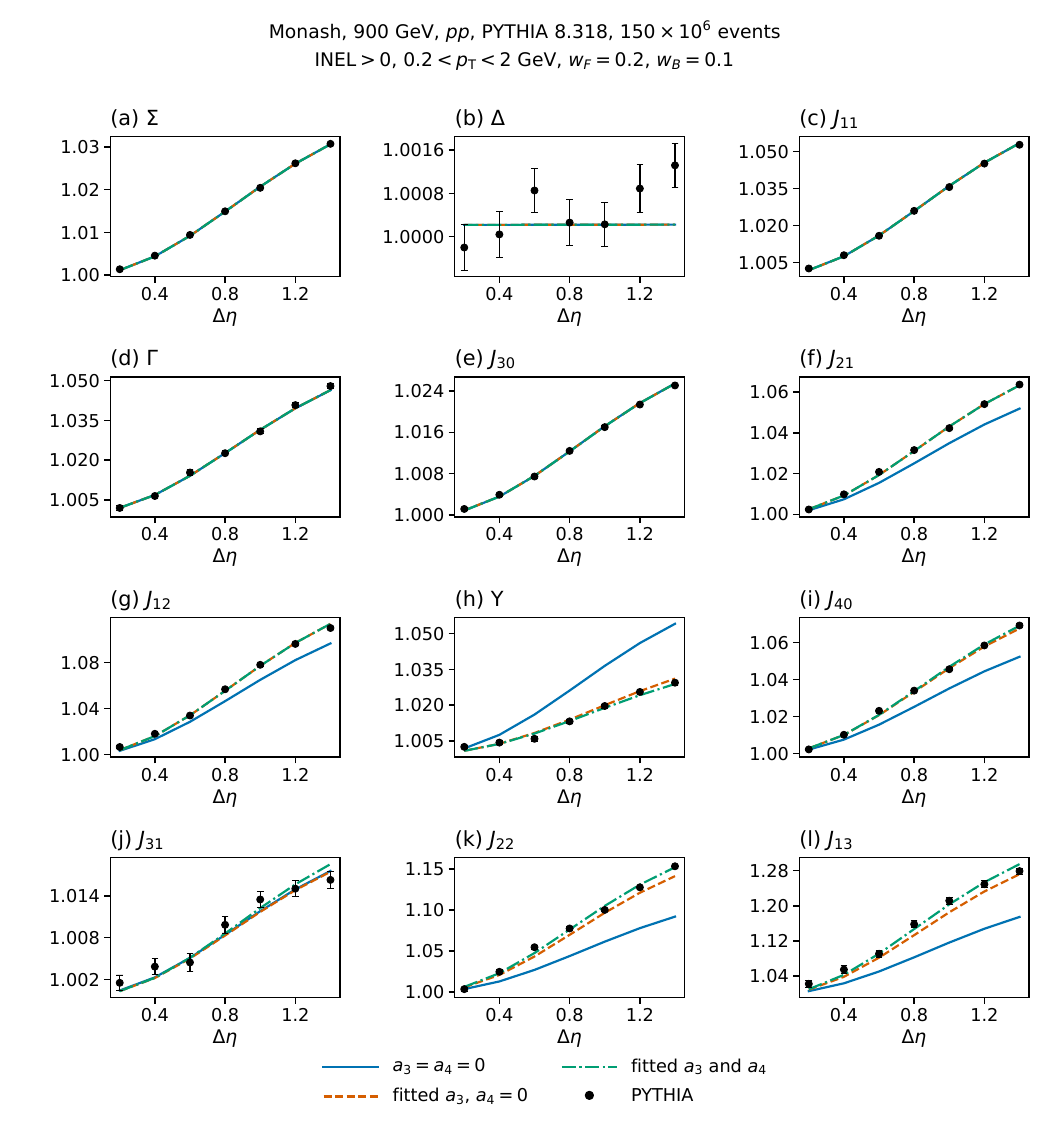}
\caption{Sequential fits with connected three- and four-particle terms (curves) compared with PYTHIA 8.318 results (points) for Monash at $\sqrt{s}=900~\mathrm{GeV}$, using 150 million generated $pp$ events. The selection is INEL$>0$ with $0.2<p_{\mathrm T}<2$ GeV and window widths $0.2/0.1$. Bars show jackknife standard errors. Lower-order parameters are held fixed and their uncertainties are not propagated into bands. Panels (a)--(l) show $\Sigma$, $\Delta$, $J_{11}$, $\Gamma$, $J_{30}$, $J_{21}$, $J_{12}$, $\Upsilon$, $J_{40}$, $J_{31}$, $J_{22}$, and $J_{13}$, respectively.}
\label{fig:connected8318_monash900}
\end{figure*}

\begin{figure*}[!t]
\centering
\includegraphics[width=\textwidth]{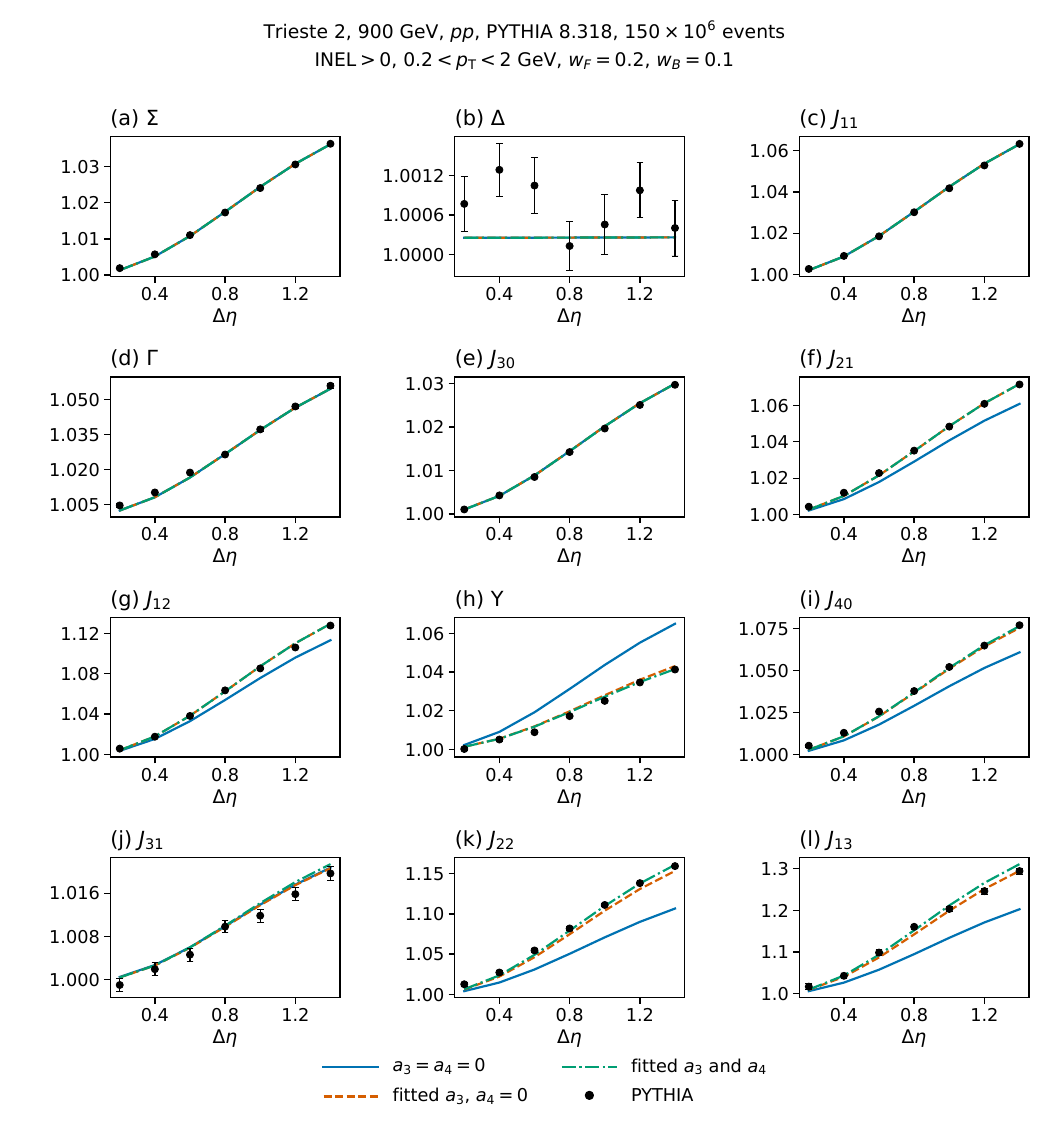}
\caption{Sequential fits with connected three- and four-particle terms (curves) compared with PYTHIA 8.318 results (points) for Trieste~2 at $\sqrt{s}=900~\mathrm{GeV}$, using 150 million generated $pp$ events. The selection is INEL$>0$ with $0.2<p_{\mathrm T}<2$ GeV and window widths $0.2/0.1$. Bars show jackknife standard errors. Lower-order parameters are held fixed and their uncertainties are not propagated into bands. Panels (a)--(l) show $\Sigma$, $\Delta$, $J_{11}$, $\Gamma$, $J_{30}$, $J_{21}$, $J_{12}$, $\Upsilon$, $J_{40}$, $J_{31}$, $J_{22}$, and $J_{13}$, respectively.}
\label{fig:connected8318_trieste900}
\end{figure*}

\begin{figure*}[!t]
\centering
\includegraphics[width=\textwidth]{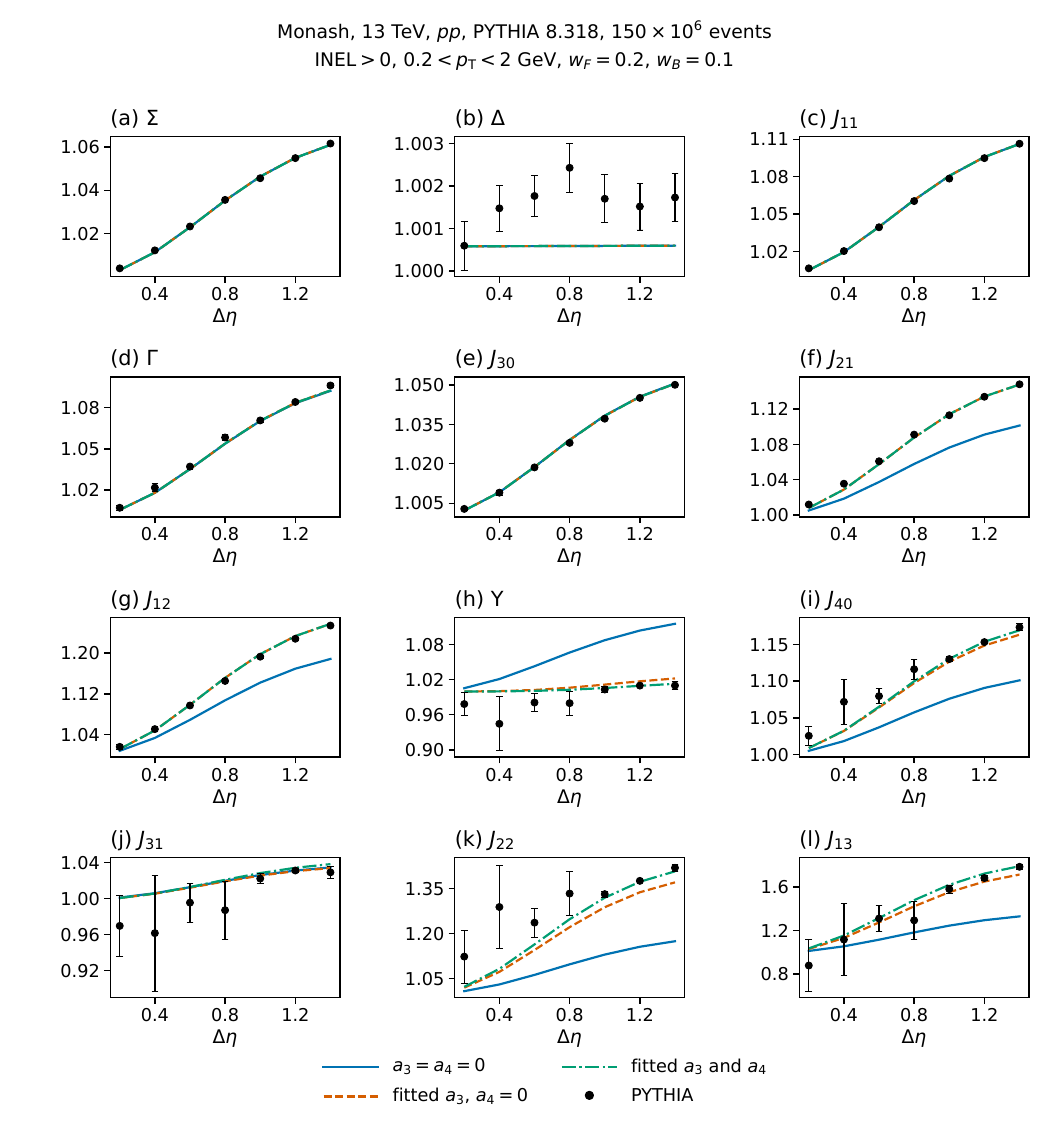}
\caption{Sequential fits with connected three- and four-particle terms (curves) compared with PYTHIA 8.318 results (points) for Monash at $\sqrt{s}=13~\mathrm{TeV}$, using 150 million generated $pp$ events. The selection is INEL$>0$ with $0.2<p_{\mathrm T}<2$ GeV and window widths $0.2/0.1$. Bars show jackknife standard errors. Lower-order parameters are held fixed and their uncertainties are not propagated into bands. Panels (a)--(l) show $\Sigma$, $\Delta$, $J_{11}$, $\Gamma$, $J_{30}$, $J_{21}$, $J_{12}$, $\Upsilon$, $J_{40}$, $J_{31}$, $J_{22}$, and $J_{13}$, respectively.}
\label{fig:connected8318_monash13000}
\end{figure*}

Figures~\ref{fig:connected8318_monash900}-- \ref{fig:connected8318_monash13000} display the resulting higher-order observables. All three comparisons use pseudorapidity windows of widths $0.2/0.1$, centered at $\pm\Delta\eta/2$, with separations up to $1.4$, INEL$>0$ event selection, and $0.2<p_{\mathrm T}<2$~GeV. Points carry jackknife standard errors. The curves are conditional central predictions, without propagated uncertainty bands from preceding fit stages.

The improvement demonstrates that the specified three- and four-particle terms account for a substantial part of the higher-order discrepancy. It does not uniquely determine the underlying connected densities: their shapes and common length are imposed, and residual discrepancies remain.

\subsection{\label{sec:pythia_alice}Application to ALICE at 13 TeV}
We combine the published ALICE pseudorapidity density at 13~TeV~\cite{ALICE2016Density13} with preliminary measurements of $\Sigma$~\cite{ErokhinIS2021} to construct conditional predictions for the higher-order observables. The density and $\Sigma$ constrain the lower-order model, while additional three- and four-particle amplitudes are specified by a transfer from the Monash calculations.

The published ALICE pseudorapidity density, measured over the full $p_{\mathrm T}$ range~\cite{ALICE2016Density13}, is converted to $0.2<p_{\mathrm T}<2$~GeV using an independent auxiliary Monash PYTHIA 8.318 sample. The paired acceptance-correction procedure is described in Appendix~\ref{app:production}.

The preliminary $\Sigma$ marker positions are extracted graphically from the presentation in Ref.~\cite{ErokhinIS2021}. Their experimental uncertainties are not extracted. These values are used without rescaling by the one-particle acceptance correction. The comparison assumes that the preliminary measurement matches the target particle and event selections. The density correction alone does not establish this matching.

\begin{figure*}[!t]
\centering
\includegraphics[width=\textwidth]{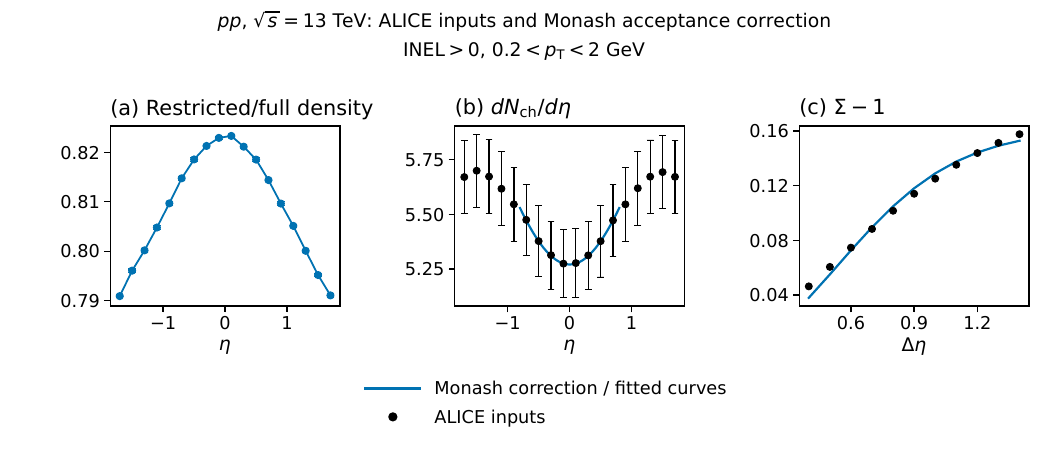}
\caption{Acceptance correction and lower-order fits for the ALICE comparison at $\sqrt{s}=13$~TeV. (a) the density correction from the full $p_{\mathrm T}$ range to $0.2<p_{\mathrm T}<2$~GeV, obtained from the auxiliary Monash PYTHIA 8.318 sample. Bars denote Monte Carlo statistical errors. (b) corrected ALICE density with scaled published errors and the fitted central-region profile. (c) Gaussian pair fit to the graphically extracted preliminary ALICE $\Sigma-1$ points. Experimental uncertainties are not included for these points.}
\label{fig:alice_density_transfer}
\end{figure*}

An unweighted linear least-squares fit to eight corrected density-bin averages in $|\eta|<0.8$ gives
\begin{equation}
 \rho_0=5.271,\qquad
 c_2=0.0834,\qquad
 c_4=-0.0109.
\end{equation}
With this profile fixed, an unweighted fit of the Gaussian pair model to 11 preliminary $\Sigma$ values at center separations $\Delta\eta=0.4,0.5,\ldots,1.4$ gives
\begin{equation}
 a_2=0.769,\qquad \ell=0.522.
\end{equation}
The root-mean-square residual in $\Sigma$ is approximately $0.004$. Figure~\ref{fig:alice_density_transfer} shows the acceptance correction and the two fits. Published density errors are scaled for display but are not used as fit weights. No experimental errors are assigned to the graphically extracted $\Sigma$ points. We therefore report these fits as descriptive parameter estimates, without statistical parameter uncertainties or a covariance-based goodness-of-fit assessment. The root-mean-square residual is not an estimate of the experimental uncertainty.

Within this parametrization, the ALICE inputs give a larger $a_2$ and a smaller $\ell$ than the Monash 13~TeV fit, for which $a_2=0.499$ and $\ell=0.651$. The differences are approximately $54\%$ and $20\%$, respectively. They should not be interpreted as a statistically quantified discrepancy: the ALICE and PYTHIA fits use different constraints and weighting procedures. In addition, $a_2=\mu_0\Lambda_0$ does not determine the normalized pair amplitude $\Lambda_0$ independently of the source normalization $\mu_0$.

To specify the higher connected terms, we transfer the ratios
\begin{equation}
 r_3=\frac{a_3}{a_2^2}\simeq9.44,
 \qquad
 r_4=\frac{a_4}{a_2^3}\simeq85.5
\end{equation}
from the conditional Monash fits at 13~TeV. These ratios determine the ALICE scenario amplitudes
\begin{equation}
\begin{aligned}
 a_3^{\mathrm{tr}}&\equiv a_{3,\mathrm{ALICE}}=r_3a_{2,\mathrm{ALICE}}^2,\\
 a_4^{\mathrm{tr}}&\equiv a_{4,\mathrm{ALICE}}=r_4a_{2,\mathrm{ALICE}}^3.
\end{aligned}
\end{equation}
The superscript $\mathrm{tr}$ denotes amplitudes determined by this transfer prescription, not fitted to higher-order ALICE measurements. The prescription is an additional model assumption, rather than a consequence of strong intensivity. Only the amplitude ratios are transferred from Monash. The connected densities $C_3$ and $C_4$ are then constructed using Eq.~\eqref{eq:connected_gaussian} with the profile and pair length fitted to the ALICE inputs.

\begin{figure*}[!t]
\centering
\includegraphics[width=\textwidth]{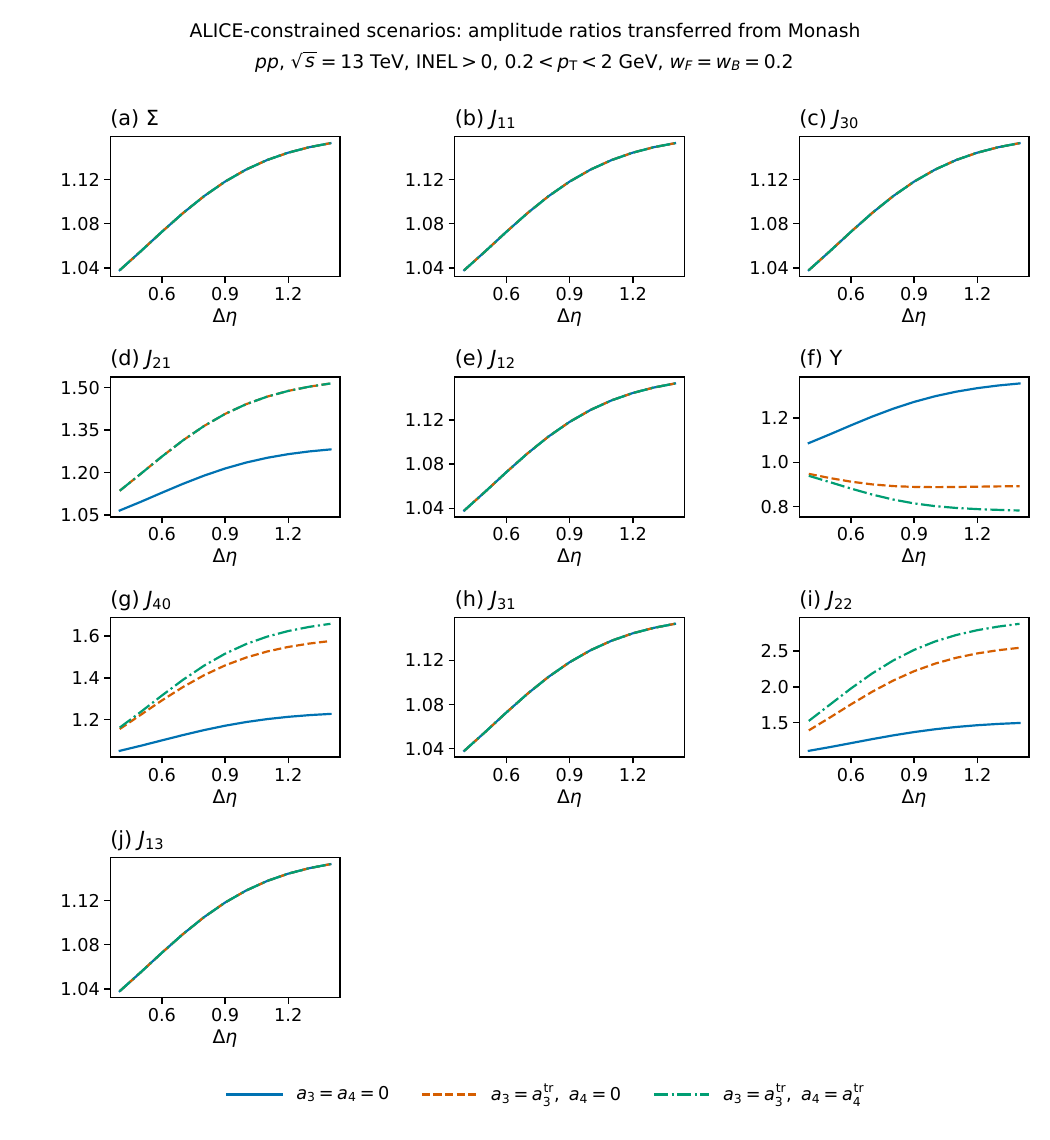}
\caption{Conditional ALICE-constrained scenarios for $pp$ collisions at $\sqrt{s}=13$ TeV with target selection INEL$>0$ and $0.2<p_{\mathrm T}<2$ GeV. Both windows have width $0.2$ and centers at $\pm\Delta\eta/2$. The scenarios share the fitted lower-order inputs. The legend specifies the higher-order amplitudes, with superscript $\mathrm{tr}$ denoting the transfer prescription of Sec.~\ref{sec:pythia_alice}. Experimental and transfer uncertainties are not propagated into bands. Panels (a)--(j) show $\Sigma$, $J_{11}$, $J_{30}$, $J_{21}$, $J_{12}$, $\Upsilon$, $J_{40}$, $J_{31}$, $J_{22}$, and $J_{13}$, respectively.}
\label{fig:alice_forecast8318}
\end{figure*}

Figure~\ref{fig:alice_forecast8318} compares three scenarios:
\begin{enumerate}
 \item the pair-only closure, $a_3=a_4=0$.
 \item a nonzero three-particle amplitude is present, $a_3=a_3^{\mathrm{tr}}$ and $a_4=0$.
 \item both higher-order amplitudes are present, $a_3=a_3^{\mathrm{tr}}$ and $a_4=a_4^{\mathrm{tr}}$.
\end{enumerate}
All three share the same fitted one- and two-particle inputs. Introducing $C_3$ changes both third- and fourth-order observables. Adding $C_4$ at fixed $C_3$ changes the fourth-order observables while leaving the third order unchanged. Even the pair-only scenario predicts nontrivial higher-order observables through the contributions of the lower-order connected densities.

These curves are conditional scenarios, not measurements of the ALICE connected densities $C_3$ and $C_4$. Their quantitative interpretation is limited by the assumed acceptance matching, graphical extraction of the preliminary points, the imperfect lower-order Monash fit at 13~TeV, and the absence of propagated experimental and transfer uncertainties. Measurements of the higher-order observables could test the transfer prescription and determine whether the three scenarios are experimentally distinguishable.

\clearpage
\section{\label{sec:discussion}Discussion and conclusions}
We have constructed a normalized hierarchy of strongly intensive observables for two particle counts in the independent identical-source model. The hierarchy contains $n$ new coordinates at order $n$. Together with the mean fraction $\alpha $, these coordinates generate all rational invariants of invertible outer composition under condition of $m\alpha \beta \ne0$. They equal one for independent particle allocation, vanish for fixed multiplicities, and remain regular at equal positive means. The construction removes the source-count distribution while retaining the invariant information in the joint cumulants.

The analytical models clarify the interpretation of the higher orders. Independent non-Poisson emission and a formal pair-only closure already generate nontrivial higher-order observables. Consequently, a departure of $J_{ij}$ from unity is not a direct measurement of a connected $(i+j)$-particle density. Identifying such contributions requires a specified lower-order baseline and an assessment of its validity, including the realizability of the assumed correlation hierarchy.

The PYTHIA study uses three samples of 150 million generated events: Monash and Trieste 2 at 900 GeV and Monash at 13 TeV, all with version 8.318. Window and event-class scans illustrate the dependence of the invariant coordinates on acceptance and particle-production dynamics. The Gaussian model of the density and pair observables gives a conditional baseline for the higher orders, but its substantial lower-order residuals at 13 TeV prevent attributing the higher-order discrepancies exclusively to $C_3$ or $C_4$. Adding specified connected terms tests a restricted hierarchy. It does not uniquely reconstruct multiparticle correlation functions.

The immediate limitations on quantitative interpretation are the lower-order model residuals and the absence of uncertainty propagation through the sequential fits. The ALICE application additionally depends on preliminary graphically extracted second-order input, assumed acceptance matching, and a generator-based acceptance correction. Experimental, fit, and transfer uncertainties must be propagated before its scenarios can be used for quantitative discrimination. A separate question for future data is whether direct higher-order measurements support the specified correlation hierarchy and the transfer of $r_3$ and $r_4$ from PYTHIA.

The strongly intensive property holds only while the selected ensemble admits independent identical sources whose emission law does not depend on the source count. Source mixtures, global conservation, and selection-induced correlations may violate this condition. Detector efficiency and finite-sample bias also require a separate analysis. Within its stated domain, the hierarchy provides a systematic means of comparing the information in successive cumulant orders while removing the dependence on the source-count distribution. Measurements of the third- and fourth-order coordinates in the same windows would test whether a hierarchy constrained at second order describes the additional invariant information.

\begin{acknowledgments}
The author acknowledges helpful conversations with Semyon Yurchenko.

The author acknowledges Saint-Petersburg State University for a research project ID: 153212394.

The author acknowledges the use of AI-based tools, including ChatGPT-6, to improve grammar, enhance clarity, and check mathematical statements. All revisions assisted by AI were carefully reviewed to ensure that the intended meaning and scientific content remain unchanged. The author takes full responsibility for the published content.
\end{acknowledgments}
\clearpage
\appendix

\section{\label{app:proof}All-order construction and completeness}
We work with formal cumulant series truncated at total degree $n\ge2$, on the domain $m\alpha\beta\ne0$. The construction below gives the invariant coordinates, their normalization, and an explicit inverse proving completeness.

\subsection{Invariant series and recursion}
Under the independent-source assumptions of Sec.~\ref{sec:definitions}, $P=Q\circ R$ with $Q'(0)=\langle S\rangle>0$. Set $D(v)=P(v,v)$ and define $h$ by
\begin{equation}
 D(h(t,u))=P(\beta t+u,-\alpha t+u).
 \label{eq:algorithm_h}
\end{equation}
Since $D'(0)=m\ne0$, the compositional inverse exists uniquely. The change of variables makes the linear part of the right-hand side $mu$, so $h=u+O((t,u)^2)$ and $h(0,u)=u$. Substitution of $P=Q\circ R$ into Eq.~\eqref{eq:algorithm_h} and cancellation of the invertible series $Q$ give
\begin{equation}
 R(h,h)=R(\beta t+u,-\alpha t+u).
\end{equation}
The weights also agree at event and source levels because $P_{10}=Q_1R_{10}$ and $P_{01}=Q_1R_{01}$. Thus $h$ is independent of the source-count law.

To compute $h$ through degree $n$, start from $h=u$ and, for $r=2,\ldots,n$, update
\begin{equation}
 h\leftarrow h+\frac{[P(\beta t+u,-\alpha t+u)-D(h)]_r}{m}.
 \label{eq:algorithm_recursion}
\end{equation}
Here $[\cdot]_r$ retains terms of total degree $r$, and the right-hand side uses the current $h$. A correction of degree $r$ changes $D(h)$ at that degree by $m$ times the correction. Contributions from the nonlinear terms of $D$ start at degree $r+1$. Each update therefore removes the current residual without changing lower orders.

\subsection{Normalization}
For the independent-particle reference, let the union multiplicity $N$ have any distribution with positive mean, and allocate each particle independently with probabilities $\alpha,\beta$. If $K$ generates the cumulants of $N$, then $P(x,y)=K(\log(\alpha e^x+\beta e^y))$ and $D(v)=K(v)$. Equation~\eqref{eq:algorithm_h} gives
\begin{equation}
 e^{h(t,u)}=e^u(\alpha e^{\beta t}+\beta e^{-\alpha t}).
 \label{eq:algorithm_independent_reference}
\end{equation}
Its derivatives are $\alpha\beta[\beta^{i-1}+(-1)^i\alpha^{i-1}]$, independently of $j$. Dividing by this reference value would be singular for $i=1$ and for odd $i$ at equal means. Instead, for $i\ge1$, $j\ge0$, and $2\le i+j\le n$, define
\begin{align}
 J_{ij}&=\frac{i!j!}{\alpha\beta}[t^iu^j]e^h+\zeta_i\Sigma,\nonumber\\
 \zeta_i&=1-\beta^{i-1}-(-1)^i\alpha^{i-1}.
 \label{eq:algorithm_output}
\end{align}
The bracket denotes an ordinary monomial coefficient. First obtain $\Sigma=2[t^2u^0]e^h/(\alpha\beta)$. Since $\zeta_2=0$, this agrees with $J_{20}=\Sigma$ without circularity. Equation~\eqref{eq:algorithm_independent_reference} yields $J_{ij}=1$. For fixed multiplicities, $h=u$ and $\Sigma=0$, hence $J_{ij}=0$. The denominator $\alpha\beta$ remains nonzero at equal positive means. The logarithm recovers $h$ from $e^h$, so exponentiation loses no information.

\subsection{\label{app:inverse}Inverse construction and completeness}
Prescribe $\alpha$ and the $J_{ij}$ through degree $n$, with $\beta=1-\alpha$ and $\alpha\beta\ne0$. Set $\Sigma=J_{20}$ and reconstruct
\begin{equation}
 E(t,u)=e^u+\alpha\beta\sum_{\substack{i\ge1,\ j\ge0\\2\le i+j\le n}}(J_{ij}-\zeta_i\Sigma)\frac{t^iu^j}{i!j!}.
 \label{eq:inverse_E}
\end{equation}
All series are truncated above degree $n$. The constant coefficient is one, so $h=\log E$ is well defined. Also $h(0,u)=u$, its linear part is $u$, and $2[t^2u^0]E/(\alpha\beta)=\Sigma$.

Choose any diagonal series $D(v)=\sum_{k=1}^n d_kv^k/k!$ with $d_1\ne0$. The inverse change of variables is $t=x-y$, $u=\alpha x+\beta y$, giving
\begin{equation}
 P(x,y)=D\!\left(h(x-y,\alpha x+\beta y)\right)
 \pmod{(x,y)^{n+1}}.
 \label{eq:inverse_P}
\end{equation}
Then $P(v,v)=D(v)$, $P_{10}=\alpha d_1$, and $P_{01}=\beta d_1$. Applying the forward construction recovers the prescribed $\alpha$ and $J_{ij}$. Both transformations are rational in their finite sets of coefficients.

For fixed $\alpha,h$, two diagonal choices give $P_2=(D_2\circ D_1^{-1})\circ P_1$. Conversely, an invertible outer composition $\widetilde P=F\circ P$ multiplies both means by $F'(0)$ and changes the diagonal to $\widetilde D=F\circ D$. It leaves $\alpha$ unchanged and gives
\begin{equation}
 \widetilde h=(F\circ D)^{-1}\circ F\circ D\circ h=h.
\end{equation}
Thus $\alpha$ and the $J_{ij}$ distinguish the outer-composition orbits. Together with the freely chosen $d_k$, they form rational coordinates on the truncated-series domain. They are therefore algebraically independent, and every rational invariant is a rational function of $\alpha$ and the $J_{ij}$.

At degree $r\ge2$, there are $r$ coefficients with $i\ge1$ and $i+j=r$. Through degree $n$, the hierarchy contains $n(n+1)/2-1$ fluctuation coordinates in addition to $\alpha$. This is a statement about formal series. Positivity and other requirements of a probability distribution impose further constraints and are not implied by the inverse construction.

\subsection{\label{app:example2}Second-order example}
The first recursion step is
\begin{align}
 h(t,u)&=u+\frac{A}{2m}t^2+\frac{B}{m}tu+O((t,u)^3),\nonumber\\
 A&=\beta^2P_{20}-2\alpha\beta P_{11}+\alpha^2P_{02},\nonumber\\
 B&=\beta(P_{20}+P_{11})-\alpha(P_{11}+P_{02}).
 \label{eq:algorithm_example}
\end{align}
Exponentiation preserves the $t^2$ and $tu$ coefficients at this order. Equation~\eqref{eq:algorithm_output} therefore gives $J_{20}=A/(m\alpha\beta)=\Sigma$ and $J_{11}=\Sigma+B/(m\alpha\beta)$, reproducing the two second-order coordinates in the main text.

\subsection{\label{app:bell}Coefficient form of the composition law}
For a zero-constant series $G(x,y)=\sum_{a+b\ge1}g_{ab}x^ay^b/(a!b!)$, the partial exponential Bell polynomials can be defined compactly by
\begin{equation}
 \mathcal B_{ij,k}(g)=\frac{i!j!}{k!}[x^iy^j]G(x,y)^k.
 \label{eq:bell_definition}
\end{equation}
The Fa\`a di Bruno composition law~\cite{DLMFFaaDiBruno,Broniowski2017} then gives
\begin{align}
 P_{ij}&=\sum_{k=1}^{i+j}Q_k\mathcal B_{ij,k}(R),\label{eq:bell_composition}\\
 i!j![t^iu^j]e^h&=\sum_{k=1}^{i+j}\mathcal B_{ij,k}(h),\label{eq:bell_exponential}
\end{align}
where the second line uses $h_{ab}=\partial_t^a\partial_u^b h(0,0)$. These finite coefficient identities provide an alternative implementation of composition and exponentiation. Invariance and completeness follow from the preceding construction.

\section{\label{app:conversions}Moment and cumulant conversions}
These identities specify the conversions used in the main text without listing every component. Write $s(i,p)$ for signed Stirling numbers of the first kind and $S(i,p)$ for Stirling numbers of the second kind, with $s(0,0)=S(0,0)=1$ and zero entries outside their usual ranges.

\subsection{Joint factorial cumulants}
Substituting $x=\log(1+z_F)$ and $y=\log(1+z_B)$ in the event cumulant series gives
\begin{equation}
 f_{ij}=\sum_{p=0}^{i}\sum_{q=0}^{j}s(i,p)s(j,q)P_{pq},
 \qquad i+j\ge1,
 \label{eq:event_factorial_conversion}
\end{equation}
where $P_{00}=0$. For example,
\begin{align}
 f_{20}&=P_{20}-P_{10},\nonumber\\
 f_{31}&=P_{31}-3P_{21}+2P_{11},\nonumber\\
 f_{22}&=P_{22}-P_{21}-P_{12}+P_{11}.
\end{align}
The inverse conversion uses $S(i,p)S(j,q)$ in place of $s(i,p)s(j,q)$.

\subsection{Inclusive densities, moments, and source cumulants}
For disjoint windows, the integrals $L_{ij}$ of Eq.~\eqref{eq:inclusive_window_integrals} are joint falling-factorial moments. Setting $L_{00}=1$, the ordinary moments are
\begin{equation}
 M_{ij}=\sum_{p=0}^{i}\sum_{q=0}^{j}S(i,p)S(j,q)L_{pq}.
 \label{eq:inclusive_to_raw_fourth}
\end{equation}
Thus $M_{10}=a$, $M_{01}=b$, and, for example, $M_{22}=L_{22}+L_{21}+L_{12}+L_{11}$. Define the truncated moment series
\begin{equation}
 \mathcal M_n(x,y)=\sum_{0\le i+j\le n}M_{ij}\frac{x^iy^j}{i!j!},
 \qquad M_{00}=1.
\end{equation}
The source cumulants through order $n$ are obtained by a finite polynomial operation:
\begin{equation}
 R_{ij}=i!j![x^iy^j]\sum_{k=1}^{n}\frac{(-1)^{k-1}}{k}
 (\mathcal M_n-1)^k,
 \label{eq:raw_to_source_fourth}
\end{equation}
for $1\le i+j\le n$. This is the truncated logarithm of the moment series. A mixed fourth-order example is
\begin{align}
 R_{22}&=M_{22}-2aM_{12}-2bM_{21}-M_{20}M_{02}\nonumber\\
 &\quad-2M_{11}^2+2a^2M_{02}+2b^2M_{20}\nonumber\\
 &\quad+8abM_{11}-6a^2b^2.
\end{align}
Setting $n=4$ gives all fourteen source cumulants needed here. When connected factorial integrals $H_{ij}$ are already known, Eq.~\eqref{eq:gauss_stirling_conversion} gives the cumulants directly and avoids the intermediate ordinary moments.
\section{\label{app:vechernin}Pair-only string model and its symmetric limit}
The pair-only model of Sec.~\ref{sec:independent_with_pair_corr} is specified through fourth order by $a=\mu_0w_F$, $b=\mu_0w_B$ and the integrals $U_F,U_B,U_{FB}$ in Eq.~\eqref{eq:exp_pair_integrals}. A compact way to generate all required source cumulants is
\begin{align}
 R(x,y)&=a z_F+b z_B+\tfrac12 U_F z_F^2\nonumber\\
 &\quad+U_{FB}z_Fz_B+\tfrac12 U_B z_B^2
 +O((x,y)^5),\nonumber\\
 z_F&=e^x-1,\qquad z_B=e^y-1.
 \label{eq:pair_generator}
\end{align}
The closure $C_3=C_4=0$ fixes the displayed terms through fourth order, while the remainder allows unrestricted higher-order connected contributions. Its derivatives reproduce Eq.~\eqref{eq:pair_ordinary_all}.

For arbitrary widths, use the general algebraic prescription in Appendix~\ref{app:gaussian_explicit} below. Insert these derivatives into Eqs.~\eqref{eq:gauss_C_def} and \eqref{eq:gauss_c_all}, then use Eqs.~\eqref{eq:gauss_named_all} and \eqref{eq:gauss_J_all}. Those algebraic formulas apply to any source cumulants and do not require the Gaussian kernel used to evaluate the integrals in Appendix~\ref{app:gaussian_explicit}. This prescription specifies all nine canonical coordinates without separate polynomial expansions. The alternative coordinate $\Upsilon$ does not add an independent invariant.

For the homogeneous source considered here, equal window widths imply $a=b$ and $U_F=U_B$. The conventional normalizations of $\Delta$ and $\Gamma$ are then undefined, as discussed after Eqs.~\eqref{eq:delta} and \eqref{eq:gamma}. All nine canonical $J_{ij}$ and $\Upsilon$ remain regular. Six canonical coordinates share the value
\begin{equation}
 J_{20}=J_{11}=J_{30}=J_{12}=J_{31}=J_{13}
 =\Sigma,
\end{equation}
where
\begin{equation}
 \Sigma=1+\frac{U_F-U_{FB}}{a}.
\end{equation}
The remaining three canonical coordinates and the alternative observable $\Upsilon$ reduce to
\begin{align}
 J_{21}
 &=1+\frac{2(U_F-U_{FB})}{a}
 -\frac{U_F^2-U_{FB}^2}{a^2},
 \nonumber\\
 J_{22}
 &=1+\frac{4(U_F-U_{FB})}{a}
 -\frac{5(U_F^2-U_{FB}^2)}{a^2}
 \nonumber\\
 &\quad+\frac{2(U_F+U_{FB})^2(U_F-U_{FB})}{a^3},
 \label{eq:pair_symmetric_basis}
\end{align}
and
\begin{align}
 J_{40}
 &=1+\frac{7(U_F-U_{FB})}{4a}
 -\frac{3(U_F^2-U_{FB}^2)}{2a^2}
 \nonumber\\
 &\quad-\frac{3(U_F+U_{FB})(U_F-U_{FB})^2}{4a^3},
 \nonumber\\
 \Upsilon
 &=1+\frac{5(U_F-U_{FB})}{3a}
 +\frac{4U_F(U_F-U_{FB})}{a^2}
 \nonumber\\
 &\quad+\frac{2(U_F-U_{FB})^2(2U_F+U_{FB})}{3a^3}.
 \label{eq:pair_symmetric_fourth}
\end{align}
\section{Production settings and auxiliary acceptance correction}
\label{app:production}
\subsection{Trieste 2 settings}
Table~\ref{tab:tunes} specifies every entry applied after \texttt{Tune:pp=14} to obtain the Trieste 2 sample. Entries that keep the baseline value are included so that the complete card is explicit.

\begin{table*}[t]
\caption{Complete Trieste 2 override card and the corresponding Monash 8.318 values. Both samples start from \texttt{Tune:pp=14}.}
\label{tab:tunes}
\small
\begin{ruledtabular}\begin{tabular}{lll}
Setting & Monash & Trieste 2\\
\texttt{StringZ:useOldAExtra} & on & off\\
\texttt{StringPT:sigma} & 0.335 & 0.335\\
\texttt{StringZ:aLund} & 0.68 & 0.36\\
\texttt{StringZ:bLund} & 0.98 & 0.56\\
\texttt{StringFlav:probQQtoQ} & 0.081 & 0.078\\
\texttt{StringFlav:probStoUD} & 0.217 & 0.2\\
\texttt{StringFlav:probQQ1toQQ0join} & \{0.5,0.7,0.9,1.\} & \{0.5,0.7,0.9,1.0\}\\
\texttt{BeamRemnants:remnantMode} & 0 & 1\\
\texttt{BeamRemnants:saturation} & 5. & 5\\
\texttt{ColourReconnection:mode} & 0 & 1\\
\texttt{ColourReconnection:timeDilationMode} & 2 & 2\\
\texttt{ColourReconnection:allowDoubleJunRem} & on & off\\
\texttt{ColourReconnection:allowJunctions} & on & on\\
\texttt{MultiPartonInteractions:pT0Ref} & 2.28 & 2.22\\
\texttt{ColourReconnection:mPseudo} & 0.3 & 0.4\\
\texttt{ColourReconnection:m0} & 0.3 & 0.5\\
\texttt{ColourReconnection:junctionCorrection} & 1.2 & 1.67\\
\texttt{ColourReconnection:timeDilationPar} & 0.18 & 0.17\\
\texttt{ClosePacking:doClosePacking} & off & on\\
\texttt{ClosePacking:PT0} & 2. & 1.75\\
\texttt{ClosePacking:enhanceStrange} & 0.08 & 0.010\\
\texttt{ClosePacking:enhancePT} & 0.5 & 0.005\\
\texttt{ClosePacking:doEnhanceDiquark} & on & off\\
\texttt{StringFragmentation:doStrangeJunctions} & off & on\\
\texttt{StringFragmentation:enhanceStrangeJunction} & 0.5 & 0.57\\
\texttt{ClosePacking:baryonSup} & 0.5 & 0.54\\
\texttt{ClosePacking:parallelBaryonSup} & 0. & 0.0\\
\end{tabular}\end{ruledtabular}
\end{table*}

\subsection{Auxiliary acceptance correction}
The published ALICE pseudorapidity density, measured over the full $p_{\mathrm T}$ range \cite{ALICE2016Density13}, is converted to the target acceptance using an auxiliary sample of one million generated Monash PYTHIA 8.318 events. Full- and restricted-$p_{\mathrm T}$ multiplicities are recorded in the same events, preserving their correlation in the acceptance correction. The sample contains 862\,339 events passing INEL$>0$. The correction uses 50 paired jackknife blocks. It is independent of the three 150-million-event samples used above.

The preliminary $\Sigma$ values are graphically extracted from the presentation in Ref.~\cite{ErokhinIS2021}. They are not rescaled by this one-particle correction, and matching the particle and event selections remains a separate assumption.
\section{Explicit window cumulants and observables for the fitted Gaussian hierarchy}
\label{app:gaussian_explicit}

This appendix evaluates the Gaussian hierarchy of Eq.~\eqref{eq:connected_gaussian} through fourth order. We consider disjoint windows $F=[\eta_F^-,\eta_F^+]$ and $B=[\eta_B^-,\eta_B^+]$. Their disjointness is required for the subsequent conversion of integrated connected factorial densities to ordinary joint cumulants.

\subsection{Window integrals of the connected densities}
\label{app:gauss_integrals}

For the one-particle density $\lambda_1(\eta)=\mu_0 f(\eta)$, with $f(\eta)=1+c_2\eta^2+c_4\eta^4$, the window means are
\begin{align}
 a=R_{10}
 &=\mu_0\left[
 \eta+\frac{c_2\eta^3}{3}+\frac{c_4\eta^5}{5}
 \right]_{\eta_F^-}^{\eta_F^+},
 \\
 b=R_{01}
 &=\mu_0\left[
 \eta+\frac{c_2\eta^3}{3}+\frac{c_4\eta^5}{5}
 \right]_{\eta_B^-}^{\eta_B^+}.
\end{align}
For nonnegative integers $i,j$ with $2\leq n=i+j\leq4$, define
\begin{equation}
 H_{ij}
 =\int_{F^i\times B^j}
 C_n(\eta_1,\ldots,\eta_n)
 \,d\eta_1\cdots d\eta_n.
 \label{eq:gauss_H}
\end{equation}
These are integrated connected factorial densities, rather than inclusive densities or ordinary cumulants. They are determined by the model parameters and introduce no additional fit parameters. Their conversion to ordinary source cumulants is given in the next subsection.

Writing $\bar\eta=n^{-1}\sum_r\eta_r$ gives $\sum_{r<s}(\eta_r-\eta_s)^2=n\sum_r(\eta_r-\bar\eta)^2$. Completing the square in $z$ then reduces each connected integral to a one-dimensional quadrature. For $\ell>0$, integration over $z$ therefore gives
\begin{equation}
 \int_{-\infty}^{\infty}
 e^{-\sum_r(\eta_r-z)^2/\ell^2}\,dz
 =\frac{\sqrt{\pi}\ell}{\sqrt n}
 e^{-\sum_{r<s}(\eta_r-\eta_s)^2/(n\ell^2)}.
 \label{eq:gauss_kernel_identity}
\end{equation}
Substituting this identity into Eq.~\eqref{eq:gauss_H} yields
\begin{align}
 H_{ij}
 &=\frac{\mu_0 a_n\sqrt n}{\sqrt{\pi}\ell}
 \int_{-\infty}^{\infty}
 F_F(z)^i F_B(z)^j\,dz,
 \label{eq:gauss_H_1d}\\
 F_W(z)
 &=\int_{\eta_W^-}^{\eta_W^+}
 f(\eta)e^{-(\eta-z)^2/\ell^2}\,d\eta,
 \qquad W\in\{F,B\}.
 \label{eq:gauss_window_function}
\end{align}

For the polynomial profile, the inner integral can be evaluated explicitly. Using $c_0=1$ in the following finite sum, the substitution $s=(\eta-z)/\ell$ gives
\begin{align}
 F_W(z)
 &=\ell\sum_{k\in\{0,2,4\}}c_k
 \sum_{r=0}^{k}\binom{k}{r}z^{k-r}\ell^r
 \nonumber\\
 &\quad\times
 \left[
 I_r\!\left(\frac{\eta_W^+-z}{\ell}\right)
 -
 I_r\!\left(\frac{\eta_W^--z}{\ell}\right)
 \right].
 \label{eq:gauss_window_explicit}
\end{align}
Here $I_r$ is an antiderivative satisfying $I_r'(t)=t^r e^{-t^2}$. A convenient choice is
\begin{align}
 I_0(t)&=\frac{\sqrt{\pi}}{2}\operatorname{erf}(t),
 \nonumber\\
 I_1(t)&=-\frac12e^{-t^2},
 \nonumber\\
 I_r(t)&=-\frac12t^{r-1}e^{-t^2}
     +\frac{r-1}{2}I_{r-2}(t),
 \qquad r=2,3,4.
 \label{eq:gauss_primitives}
\end{align}
Only differences of these antiderivatives enter Eq.~\eqref{eq:gauss_window_explicit}, so their additive constants are immaterial.

Equations~\eqref{eq:gauss_H_1d}, \eqref{eq:gauss_window_explicit}, and \eqref{eq:gauss_primitives} reduce every required connected integral through fourth order to a one-dimensional quadrature for the finite windows and the fitted nonuniform profile. The reduction is exact. Only the remaining integration over $z$ requires numerical evaluation.
\subsection{\label{app:gauss_cumulants}All fourteen ordinary source cumulants}
For disjoint windows, the conversion from integrated connected factorial densities to ordinary cumulants follows from
\begin{equation}
 R(x,y)=K(e^x-1,e^y-1),
 \label{eq:gauss_factorial_composition}
\end{equation}
where $K$ is the joint factorial cumulant generating function of the source multiplicities. Through fourth order,
\begin{align}
 K(z_F,z_B)
 &=az_F+bz_B+\sum_{\substack{p,q\geq0\\2\leq p+q\leq4}}H_{pq}\frac{z_F^p z_B^q}{p!q!}\nonumber\\
 &\quad+O\bigl((z_F,z_B)^5\bigr).
\end{align}
Terms of higher total degree cannot contribute to ordinary cumulants through fourth order.

For a compact expression, extend the notation by $H_{10}=a$, $H_{01}=b$, and $H_{00}=0$. Differentiating Eq.~\eqref{eq:gauss_factorial_composition} gives
\begin{equation}
 R_{ij}
 =\sum_{p=0}^{i}\sum_{q=0}^{j}
 \begin{Bmatrix}i\\p\end{Bmatrix}
 \begin{Bmatrix}j\\q\end{Bmatrix}
 H_{pq},
 \qquad 1\leq i+j\leq4,
 \label{eq:gauss_stirling_conversion}
\end{equation}
where the braces denote Stirling numbers of the second kind, with $\left\{\begin{smallmatrix}0\\0\end{smallmatrix}\right\}=1$. These coefficients count partitions of the labeled derivatives associated with each window into nonempty blocks. They are a specialization of the Bell-polynomial composition law discussed in Appendix~\ref{app:bell}.

The complete list of fourteen ordinary source cumulants is
\begin{align}
 R_{10}&=a,
 & R_{01}&=b,
 \nonumber\\
 R_{20}&=a+H_{20},
 & R_{02}&=b+H_{02},
 \nonumber\\
 R_{11}&=H_{11},
 \nonumber\\
 R_{30}&=a+3H_{20}+H_{30},
 \nonumber\\
 R_{03}&=b+3H_{02}+H_{03},
 \nonumber\\
 R_{21}&=H_{11}+H_{21},
 & R_{12}&=H_{11}+H_{12},
 \nonumber\\
 R_{40}&=a+7H_{20}+6H_{30}+H_{40},
 \nonumber\\
 R_{04}&=b+7H_{02}+6H_{03}+H_{04},
 \nonumber\\
 R_{31}&=H_{11}+3H_{21}+H_{31},
 \nonumber\\
 R_{22}&=H_{11}+H_{21}+H_{12}+H_{22},
 \nonumber\\
 R_{13}&=H_{11}+3H_{12}+H_{13}.
 \label{eq:gauss_R_all}
\end{align}
The terms proportional to $H_{30}$, $H_{21}$, $H_{12}$, and $H_{03}$ show why $C_3$ affects fourth-order predictions even when $C_4=0$. Conversely, $C_4$ cannot enter cumulants below fourth order. The conversion is linear in the connected integrals. Products of lower-order correlations arise only when the cumulants are combined into the nonlinear observables.

\subsection{\label{app:gauss_observables}Explicit finite formulas for every strongly intensive coordinate}
Here the general construction is applied to the single-source cumulant series $R$, so $m=a+b$ is the mean multiplicity of one source in the two windows. Set $\alpha=a/m$ and $\beta=b/m$. Define the normalized directional cumulants
\begin{equation}
 \mathcal K_{ij}
 =\frac{1}{m}
 \left.
 \partial_t^i\partial_u^j
 R(\beta t+u,-\alpha t+u)
 \right|_{t=u=0}.
 \label{eq:gauss_C_def}
\end{equation}
They are the joint cumulants of the source imbalance $\beta\mu_F-\alpha\mu_B$ and total multiplicity $\mu_F+\mu_B$, divided by $m$. They introduce no additional fit parameters and are distinct from both the profile coefficients $c_2,c_4$ and the connected densities $C_n$. Their dependence on the latter is determined by the window integrals and Eq.~\eqref{eq:gauss_R_all}.

All required directional cumulants follow from the finite binomial sum
\begin{equation}
 m\mathcal K_{ij}=\sum_{p=0}^{i}\sum_{q=0}^{j}\binom{i}{p}\binom{j}{q}
 \beta^p(-\alpha)^{i-p}R_{p+q,i+j-p-q}.
 \label{eq:gauss_c_all}
\end{equation}
For example, $m\mathcal K_{20}=\beta^2R_{20}-2\alpha\beta R_{11}+\alpha^2R_{02}$. This formula is simply the expansion of $(\beta\partial_x-\alpha\partial_y)^i(\partial_x+\partial_y)^jR$ at the origin.

The conventional second- and third-order observables $\Delta$ and $\Gamma$ are
\begin{align}
 \Delta
 &=\frac{\beta^2R_{20}-\alpha^2R_{02}}
 {m\alpha\beta(\beta-\alpha)},
 \\
 \Gamma
 &=\frac{\mathcal K_{30}}
 {\alpha\beta(\beta-\alpha)}.
 \label{eq:gauss_conv_pair}
\end{align}
These normalizations require $\alpha\ne\beta$. The alternative fourth-order coordinate is
\begin{align}
 \Upsilon
 &=\left(
 \mathcal K_{02}
 +\frac{(\beta-\alpha)\mathcal K_{11}}{\alpha\beta}
 \right)
 \left(\frac{\mathcal K_{20}}{\alpha\beta}\right)^2
 \nonumber\\
 &\quad+\left(\frac{1}{3\alpha\beta}-1\right)
 \left(\frac{\mathcal K_{20}}{\alpha\beta}\right)^3
 -\frac{\mathcal K_{40}}{3\alpha^2\beta^2}.
 \label{eq:gauss_named_all}
\end{align}

Evaluating Eq.~\eqref{eq:algorithm_output} through fourth order gives the nine canonical coordinates:
\begin{align}
 J_{20}=\Sigma
 &=\frac{\mathcal K_{20}}{\alpha\beta},
 \nonumber\\
 J_{11}
 &=\frac{\mathcal K_{20}+\mathcal K_{11}}{\alpha\beta},
 \nonumber\\
 J_{30}
 &=\frac{\mathcal K_{30}+2\alpha\mathcal K_{20}}
 {\alpha\beta},
 \nonumber\\
 J_{21}
 &=\frac{\mathcal K_{21}
 +(1-\mathcal K_{02})\mathcal K_{20}}{\alpha\beta},
 \nonumber\\
 J_{12}
 &=\frac{\mathcal K_{12}
 +2(1-\mathcal K_{02})\mathcal K_{11}
 +\mathcal K_{20}}{\alpha\beta},
 \nonumber\\
 J_{40}
 &=\frac{\mathcal K_{40}
 +3(1-\mathcal K_{02})\mathcal K_{20}^2}
 {\alpha\beta}
 +3\mathcal K_{20},
 \nonumber\\
 J_{31}
 &=\frac{1}{\alpha\beta}
 \bigl[
 \mathcal K_{31}+2\alpha\mathcal K_{20}
 \nonumber\\
 &\qquad+(1-\mathcal K_{02})
 (\mathcal K_{30}+3\mathcal K_{20}\mathcal K_{11})
 \bigr],
 \nonumber\\
 J_{22}
 &=\frac{1}{\alpha\beta}
 \bigl[
 \mathcal K_{22}
 +2(1-\mathcal K_{02})(\mathcal K_{21}+\mathcal K_{11}^2)
 \nonumber\\
 &\qquad+
 (1-2\mathcal K_{02}+2\mathcal K_{02}^2-\mathcal K_{03})
 \mathcal K_{20}
 \bigr],
 \nonumber\\
 J_{13}
 &=\frac{1}{\alpha\beta}
 \bigl[
 \mathcal K_{13}
 +3(1-\mathcal K_{02})\mathcal K_{12}
 +\mathcal K_{20}
 \nonumber\\
 &\qquad+
 3(1-2\mathcal K_{02}+2\mathcal K_{02}^2-\mathcal K_{03})
 \mathcal K_{11}
 \bigr].
 \label{eq:gauss_J_all}
\end{align}

Equations~\eqref{eq:gauss_H_1d}, \eqref{eq:gauss_R_all}, and \eqref{eq:gauss_c_all}, together with Eqs.~\eqref{eq:gauss_named_all} and \eqref{eq:gauss_J_all} provide an explicit finite evaluation of the observables for the fitted Gaussian hierarchy. No numerical differentiation or unspecified higher-order cumulants are required.

Setting $a_3=a_4=0$ recovers the pair-only expressions in terms of $a$, $b$, $U_F$, $U_B$, and $U_{FB}$, with
\begin{equation}
 U_F=H_{20},\qquad
 U_B=H_{02},\qquad
 U_{FB}=H_{11}.
\end{equation}
The integrals are evaluated here with the Gaussian kernel and fitted nonuniform profile, rather than the homogeneous exponential model of Appendix~\ref{app:vechernin}.

Setting $a_2=a_3=a_4=0$ gives independent Poisson emission into the two windows. The reference normalization of Eq.~\eqref{eq:algorithm_output} then yields $\Sigma=J_{ij}=1$, and $\Upsilon=1$. The conventional $\Delta$ and $\Gamma$ also equal one when $\alpha\ne\beta$. For equal positive means, all nine canonical $J_{ij}$ and $\Upsilon$ remain regular, whereas the conventional normalizations of $\Delta$ and $\Gamma$ are undefined.

At fixed $a_n$, profile coefficients, and $\ell$, the common factor $\mu_0$ cancels between the source cumulants and $m$ in Eq.~\eqref{eq:gauss_C_def}. Under the independent-source assumptions of Sec.~\ref{sec:definitions}, fluctuations of the number of sources also leave the observables unchanged, by the outer-composition cancellation established in Appendix~\ref{app:proof}.
\clearpage
\bibliography{apssamp}
\end{document}